\documentclass[aps,prb,reprint,superscriptaddress,floatfix]{revtex4-2}
\usepackage[utf8]{inputenc}
\usepackage{hyperref}           %For bibliography
\usepackage{amsmath}
\usepackage{marvosym}
\usepackage{amssymb,stmaryrd}   %For mathematical equations
\usepackage{bbm}
\usepackage{babel}
\usepackage{comment}
\usepackage{bm}
\usepackage{color}
\usepackage{natbib}
\usepackage{graphicx,subfigure} %For figures
\usepackage{epstopdf}
\usepackage{float}
\graphicspath{ {./figures/} }
\usepackage{dcolumn}    %Align table columns on the decimal point
\usepackage{wrapfig}
\usepackage{mathtools}
\usepackage{xspace}
\usepackage{refstyle}   %For bibliography
\usepackage{tikz}  
\usetikzlibrary{arrows,shapes,trees,positioning}  
\usepackage{siunitx}

\usepackage{appendix}
\usepackage{textcomp}
\usepackage{booktabs}
\usepackage[mathlines]{lineno}% Enable numbering of text and display math
\usepackage[normalem]{ulem}
\usepackage[ignoreunlbld,norefs,nocites]{refcheck}
\usepackage{multirow}
\hypersetup{
	colorlinks=true,
	linkcolor=blue,
	citecolor=blue,
	urlcolor=blue,
}

\definecolor{darkgreen}{rgb}{0, 0.8, 0}

\begin{document}

\title{Transition-Metal-Induced Uniaxial Anisotropy in YFe$_3$}

\author{M. Nur Hasan} 
\thanks{Contact author: \href{mailto:nur.hasan@physics.uu.se}{nur.hasan@physics.uu.se}}
\affiliation{Department of Physics and Astronomy, Uppsala University, Box 524, SE-751 20 Uppsala, Sweden}
\author{Bruno L. Fertig}
\affiliation{Institute for Metallic Materials, Leibniz IFW Dresden, 01069 Dresden, Germany}
\author{Santa Pile}
\affiliation{Center for Modeling and Simulation, University for Continuing Education Krems, Austria}
\affiliation{Solid State Physics Department, Johannes Kepler University Linz, Austria}
\author{Rafael M. Vieira}
\affiliation{Department of Physics and Astronomy, Uppsala University, Box 524, SE-751 20 Uppsala, Sweden}
\affiliation{Department of Engineering Sciences and Mathematics, Lule{\aa} University of Technology, Sweden}
\author{Alena Vishina}
\affiliation{Department of Physics and Astronomy, Uppsala University, Box 524, SE-751 20 Uppsala, Sweden}
\author{Alexander Kovacs}
\affiliation{Center for Modeling and Simulation, University for Continuing Education Krems, Austria}
\author{Gotthard Rieger}
\affiliation{Siemens AG, Munich, Germany}
\author{Stefan Denneler}
\affiliation{Siemens AG, Munich, Germany}
\author{Samuel J. R. Holt}
\affiliation{Max Planck Institute for the Structure and Dynamics, Hamburg, Germany}
\affiliation{Center for Free-Electron Laser Science (CFEL), Hamburg, Germany}
\author{Martin Lang}
\affiliation{Max Planck Institute for the Structure and Dynamics, Hamburg, Germany}
\affiliation{Center for Free-Electron Laser Science (CFEL), Hamburg, Germany}
\author{Swapneel Pathak}
\affiliation{Max Planck Institute for the Structure and Dynamics, Hamburg, Germany}
\affiliation{Center for Free-Electron Laser Science (CFEL), Hamburg, Germany}
\author{Andrea Petrocchi}
\affiliation{Max Planck Institute for the Structure and Dynamics, Hamburg, Germany}
\affiliation{Center for Free-Electron Laser Science (CFEL), Hamburg, Germany}
\author{Hans Fangohr}
\affiliation{Max Planck Institute for the Structure and Dynamics, Hamburg, Germany}
\affiliation{Center for Free-Electron Laser Science (CFEL), Hamburg, Germany}
\affiliation{University of Southampton, Southampton, United Kingdom}
\author{Thomas George Woodcock}
\affiliation{Institute for Metallic Materials, Leibniz IFW Dresden, 01069 Dresden, Germany}
\author{Thomas Schrefl}
\affiliation{Center for Modeling and Simulation, University for Continuing Education Krems, Austria}
\author{Heike C. Herper}
%\email {heike.herper@physics.uu.se}
\affiliation{Department of Physics and Astronomy, Uppsala University, Box 524, SE-751 20 Uppsala, Sweden}

\date{\today} 

%%%%%%%%%%%%%%%%%%%%%%%%%%%%%%%%%%%%%%%%%%%%%%%%%%%%%%%%%%%%%%%%%%%%%%%%%%%%%%%%%%%%%%%
%\section{Abstract}
\begin{abstract}
Achieving robust uniaxial magnetocrystalline anisotropy while preserving high magnetization and thermal stability remains a central challenge in the development of Fe-rich permanent magnets. YFe$_3$ is a promising candidate owing to its high saturation magnetization and Curie temperature, but its intrinsic easy-plane anisotropy limits its potential for permanent-magnet applications. Here, we show that selective transition-metal substitution provides an effective means of driving an easy-plane-to-easy-axis transition in YFe$_3$. Systematic first-principles screening identifies V, Nb, and Ta as particularly effective dopants, with YFe$_{2.33}$Ta$_{0.67}$ exhibiting a large uniaxial magnetocrystalline anisotropy energy of 2.3~MJ/m$^{3}$. Calculated exchange interactions and atomistic Monte Carlo simulations show that the enhanced anisotropy is achieved while preserving robust ferromagnetic order and high thermal stability. Micromagnetic simulations further reveal that YFe$_{2.33}$Ta$_{0.67}$ exhibits the highest coercivity and strongest resistance to irreversible demagnetization, whereas YFe$_{2.33}$V$_{0.67}$ achieves a comparable maximum energy product at 300~K. Experimental characterization of V-alloyed YFe$_3$ shows that the YFe$_3$ phase can be retained at low nominal V concentrations (1–3 at.\%) under the employed synthesis and heat-treatment conditions, whereas higher V contents lead to increasingly multiphase samples. Concentration-dependent MAE calculations reveal a sign reversal between approximately 11 and 14~at.\% V, marking the transition from easy-plane to uniaxial easy-axis anisotropy. Consistent with this concentration dependence, electron backscatter diffraction and Kerr microscopy reveal in-plane magnetic domain structures in the experimentally accessible low-V regime. Together, these results suggest that Fe-site transition-metal substitution offers a promising approach to tuning the magnetocrystalline anisotropy of YFe$_3$ while retaining high Curie temperatures, with the potential to promote uniaxial easy-axis anisotropy.
\end{abstract}

\maketitle

\section{Introduction}
High-performance permanent magnets require the simultaneous realization of large uniaxial magnetocrystalline anisotropy (MCA), high saturation magnetization ($M_\mathrm{s}$), and robust thermal stability (e.g. a high Curie temperature). These intrinsic properties set the fundamental magnetic energy scales that determine the coercivity, maximum energy product $[(BH)_\mathrm{max}]$, and thermal behaviour of a magnetic material \cite{CoeyBook2010}. In particular, MCA originates from the interplay between spin--orbit coupling and the crystal-field environment, establishing an energetic preference for magnetization along specific crystallographic directions. A large uniaxial anisotropy stabilizes the magnetization against reversal and gives rise to a strong anisotropy field, both of which are essential for achieving high coercivity and a large maximum energy product \cite{Gutfleisch2011,Coey2011}. However, the experimentally realized coercivity and energy product are also strongly influenced by microstructural features such as grain boundaries, crystallographic defects, and domain-wall pinning. Consequently, the development of high-performance permanent magnets requires the optimization of both intrinsic electronic-structure-derived magnetic properties and extrinsic microstructural characteristics \cite{coeyPerspectiveProspectsRare2020,fischbacher2018}

The rapid expansion of electrification technologies, renewable-energy systems, and advanced information-storage devices has intensified the demand for high-performance permanent magnets \cite{mccallum2014practical}. Rare-earth magnets, particularly Nd--Fe--B and Sm--Co, remain the benchmark materials owing to their unique combination of large uniaxial MCA, high saturation magnetization, and excellent thermal stability \cite{Gutfleisch2011,Sagawa1984,RevModPhys.63.819,HONO2012530}. However, concerns over the cost, supply security, and environmental footprint of critical rare-earth elements have stimulated intense research into rare-earth-lean and rare-earth-free alternatives \cite{mccallum2014practical,NAKAMURA2018273,lewis2013}. Developing such materials while retaining the magnetic performance required for permanent-magnet applications remains a major challenge.

Among the candidate rare-earth-lean systems, Fe-rich intermetallic compounds have attracted considerable attention owing to their large magnetic moments, strong exchange interactions, and high Curie temperatures arising from the Fe sublattice. Representative examples include Fe$_{16}$N$_2$ \cite{Metzger1994,Coey1994}, Fe$_2$P \cite{JPSJ.43.41,PhysRevB.86.085125,PhysRevB.82.085103}, L1$_0$-ordered FeNi \cite{Miura_2013}, Heusler compounds \cite{GRAF20111,PhysRevB.107.174402,jami2023,faleev2017,meng2019,gasi2013,PhysRevB.98.014411,PhysRevMaterials.6.025402}, and Mn-based magnetic materials \cite{Cui2014,Cui_2014,Rao_2013,PhysRevB.90.054404,JIAN2015524,ENER2015265}. Extensive efforts have focused on enhancing magnetocrystalline anisotropy through lattice-strain engineering, chemical substitution, interstitial alloying, and crystal-structure design \cite{Warnicke_2007,PhysRevB.86.085125,PhysRevB.80.064415,PhysRevB.86.174430,PhysRevLett.96.037205,Neise2011,PhysRevLett.93.027203,Reichel2014,PhysRevB.89.144403}. Nevertheless, the realization of Fe-rich materials that combine large uniaxial anisotropy, high magnetization, and robust thermal stability without relying on critical rare-earth elements remains an important goal in the search for next-generation permanent magnets.

The Y--Fe intermetallic family, including Y$_6$Fe$_{23}$, Y$_2$Fe$_{17}$, YFe$_3$, and YFe$_2$, has attracted considerable attention owing to its rich structural diversity and favorable magnetic properties \cite{buschow1976,coehoorn1989,sabiryanov1998}. Y$_6$Fe$_{23}$ and YFe$_2$ crystallize in cubic structures and exhibit relatively soft magnetic behavior, whereas Y$_2$Fe$_{17}$ and YFe$_3$ occur in both hexagonal and rhombohedral crystal structures. Although Y$_2$Fe$_{17}$ possesses a high Fe content, its relatively low Curie temperature of 300--325~K limits its applicability for permanent-magnet applications \cite{VORONIN200182,Nikitin_jap}. In contrast, YFe$_3$ combines a high Curie temperature of 539--569~K \cite{Oesterreicher,buschow1976hydrogen,bolyachkin2017,givord1971}, comparable to that of the benchmark Nd$_2$Fe$_{14}$B permanent magnet ($T_\mathrm{C}\approx585$~K) \cite{Sagawa1984,RevModPhys.63.819}, with a large spontaneous magnetic polarization of approximately 0.94~T \cite{givord1971,buschow1976hydrogen}. These properties make YFe$_3$ an attractive platform for the development of rare-earth-lean permanent magnets. However, YFe$_3$ exhibits pronounced easy-plane anisotropy, characterized by a first anisotropy constant of $K_1=-1.1$~MJ/m$^{3}$ at 2~K \cite{bolyachkin2017}, which prevents its use as a permanent magnet. Developing an effective strategy to transform this intrinsic easy-plane anisotropy into robust uniaxial anisotropy while preserving the large magnetization and high Curie temperature therefore remains a key challenge.

Chemical substitution offers a promising route to overcome this limitation by modifying the local electronic structure and crystal-field environment, thereby enabling control of magnetocrystalline anisotropy \cite{PhysRevB.92.174413,PhysRevB.99.024421}. Here, we combine first-principles calculations, atomistic Monte Carlo simulations, micromagnetic modelling, and experimental characterization to establish a multiscale framework for anisotropy engineering in YFe$_3$. By systematically screening transition-metal dopants at all crystallographically distinct Fe and Y sites, we identify V, Nb, and Ta as effective elements for inducing an easy-plane-to-easy-axis transition. We further reveal how chemical substitution modifies the electronic structure, magnetic exchange interactions, and finite-temperature magnetic properties, and assess the impact of these changes on coercivity and energy-product performance. Finally, experimental investigations of V-doped YFe$_3$
probe the low-doping regime accessible through arc melting
and subsequent heat treatment, highlighting the challenges
of achieving higher substitution levels using this
synthesis approach.

%TC:ignore
\section{Methodology}
The methodology combines first-principles and finite-temperature magnetic simulations, micromagnetic simulations, and experimental characterization. Details are provided in Secs.~\ref{ab-initio}, \ref{mm}, and \ref{exp-details}, respectively.

\subsection{Ab initio and finite-temperature simulations}
\label{ab-initio}
Spin-polarized density functional theory (DFT) calculations were performed using the Vienna \textit{ab initio} Simulation Package (VASP)~\cite{KRESSE199} and the FP-LMTO-based code RSPt~\cite{wills2010full} to investigate YFe$_3$ and its doped configurations. The lattice parameters and formation energies were computed using the Perdew--Burke--Ernzerhof (PBE) generalized-gradient approximation (GGA) exchange-correlation functional \cite{PhysRevLett.77.3865} and the projected augmented-wave (PAW) pseudopotentials \cite{PhysRevB.59.1758} as implemented in VASP. We set the plane-wave expansion cutoff to \SI{520}{eV} and the energy convergence criterion to \SI{e-5}{eV} for electronic iterations. For structural optimization, we used the conjugate gradient algorithm, continuing until all Hellmann--Feynman forces were below \SI{0.01}{eV/\angstrom}. The formation energies of all doped configurations were calculated to identify energetically favorable substitutional structures. The energetically favorable configurations were subsequently selected for intrinsic magnetic property calculations using RSPt~\cite{wills2010full}.

To investigate how dopants replace Fe, we modeled the chemical disorder using special quasirandom structures (SQS)~\cite{PhysRevLett.65.353}. The SQS were generated using the \textit{mcsqs} tool \cite{VANDEWALLE201313} from the Alloy Theoretic Automated Toolkit (ATAT) package~\cite{van2002automating,VANDEWALLE2002539,VANDEWALLE2009266}, with the Pymatgen package~\cite{ONG2013314} as an interface (see Supplementary Sec. I.A for details).
We generated SQS for all possible combinations  of occupancy  in Fe sublattices: one set of structures with 36-atom supercells (minimal size to accommodate the doping concentration) for computationally efficient first-principles structural relaxation calculations (performed with VASP), and another with 72-atom supercells to better capture disorder, serving as a control set to verify the robustness of the trends observed in the smaller SQS. For the fully disordered case, with each Fe site having an equal probability of being occupied by the dopant, the smallest SQS is a 108-atom supercell, which was compared against the control SQS with 324 atoms. The control involved comparing the total energies of both SQS sets, estimated from the relaxed supercells using Mattersim (5M)~\cite{yang2024mattersim}, a universal machine learning interatomic potential (UMLIP), in conjunction with the Atomic Simulation Environment (ASE) python package~\cite{HjorthLarsen_2017}. The supercells were relaxed under hydrostatic strain constraints using the ASE symmetrizer FrechetCellFilter and the BFGS optimizer, with a force convergence criterion of 0.0002 eV/\AA.

The magnetocrystalline anisotropy energy (MAE) was calculated using RSPt~\cite{wills2010full} with a converged dense $32 \times 32 \times 18$ $k$-point mesh. MAE is a key intrinsic property that determines the orientation of the magnetic easy axis and provides the anisotropy field required for magnetic stability and high coercivity in permanent magnets. It arises from the interplay between SOC and the crystal-field environment \cite{skomski2008simple}. Both the magnetic force theorem~\cite{WANG1996337} and the total-energy difference method were employed, with the MAE ($K_u$) defined as

\begin{equation}
K_u = E_{[100]} - E_{[001]},
\label{eq:mae}
\end{equation}

where $E_{[100]}$ and $E_{[001]}$ denote the energies (or the band energies within the magnetic force theorem) for magnetization oriented along the [100] and [001] crystallographic directions, respectively. Within the magnetic force theorem, the band energies were obtained non-self-consistently by including spin--orbit coupling (SOC) on the converged self-consistent scalar-relativistic charge density.

The Heisenberg exchange interactions ($J_{ij}$) were calculated using the Liechtenstein--Katsnelson--Antropov--Gubanov (LKAG) formalism \cite{LIECHTENSTEIN198765}, as implemented in the RSPt code~\cite{PhysRevB.91.125133}. Within the LKAG formalism, the exchange interaction represents the change in electronic energy associated with infinitesimal rotations of two magnetic moments and is given by

\begin{equation}
J_{ij}=\frac{1}{4\pi}\,\mathrm{Im}
\int_{-\infty}^{E_{\mathrm{F}}}
\mathrm{Tr}
\left[
\hat{\Delta}_{i}
\hat{G}^{\uparrow}_{ij}(\varepsilon)
\hat{\Delta}_{j}
\hat{G}^{\downarrow}_{ji}(\varepsilon)
\right]
d\varepsilon,
\label{eq:lkag}
\end{equation}

where $E_{\mathrm{F}}$ is the Fermi energy, $\hat{\Delta}_{i}$ is the onsite exchange-splitting potential, and $\hat{G}^{\uparrow,\downarrow}_{ij}$ are the spin-dependent intersite Green's functions. The calculated exchange interactions were
subsequently mapped onto the classical
Heisenberg Hamiltonian,

\begin{equation}
\mathcal{H}_{\mathrm{}}
=
-\sum_{i\neq j}
J_{ij}\,\mathbf{e}_i\cdot\mathbf{e}_j,
\label{eq:heisenberg}
\end{equation}

where $J_{ij}$ denotes the pairwise exchange interaction between magnetic sites $i$ and $j$, and $\mathbf{e}_i$ is the classical unit vector along the direction of the local magnetic moment at site $i$.

An accurate description of finite-temperature magnetic behavior requires the inclusion of long-range exchange interactions. Accordingly, all pairwise exchange interactions $J_{ij}$ between a central site $i$ and neighboring sites $j$ within a radial cutoff of $3.0a$ were included in the spin Hamiltonian, where $a$ denotes the basal-plane lattice
parameter of the hexagonal unit cell. As a convergence check, the exchange interactions were also evaluated up to $4.0a$ for one representative composition, yielding no appreciable change in the calculated magnetic properties. Since the exchange interactions beyond $3.0a$ were negligible and the structural changes induced by the substitutions were small, a $3.0a$ cutoff was consistently used for the remaining compositions.

The finite-temperature magnetic properties were investigated using the Metropolis Monte Carlo algorithm implemented in the Uppsala Atomistic Spin Dynamics (UppASD) package~\cite{eriksson2017}. Simulations were performed on a $32 \times 32 \times 32$ supercell, and statistical averaging was conducted over three independent ensembles to mitigate uncertainty. For each temperature, an initial $5\times 10^4$ MC steps were used for thermal equilibration, followed by a subsequent measurement phase with $5\times 10^4$ steps for sampling thermodynamic observables. The Curie temperature was determined from heating and cooling simulations. A temperature increment of \SI{20}{K} was used over most of the investigated temperature range, with finer temperature sampling near the magnetic phase transition to accurately resolve the Curie temperature and in the low-temperature regime where required.
%%%%%%%%%%%%%%%%%%%%%%%%%%%%%%%%%%%%
\begin{figure*}[t!]
	\centering
	\includegraphics[width=0.9\textwidth]{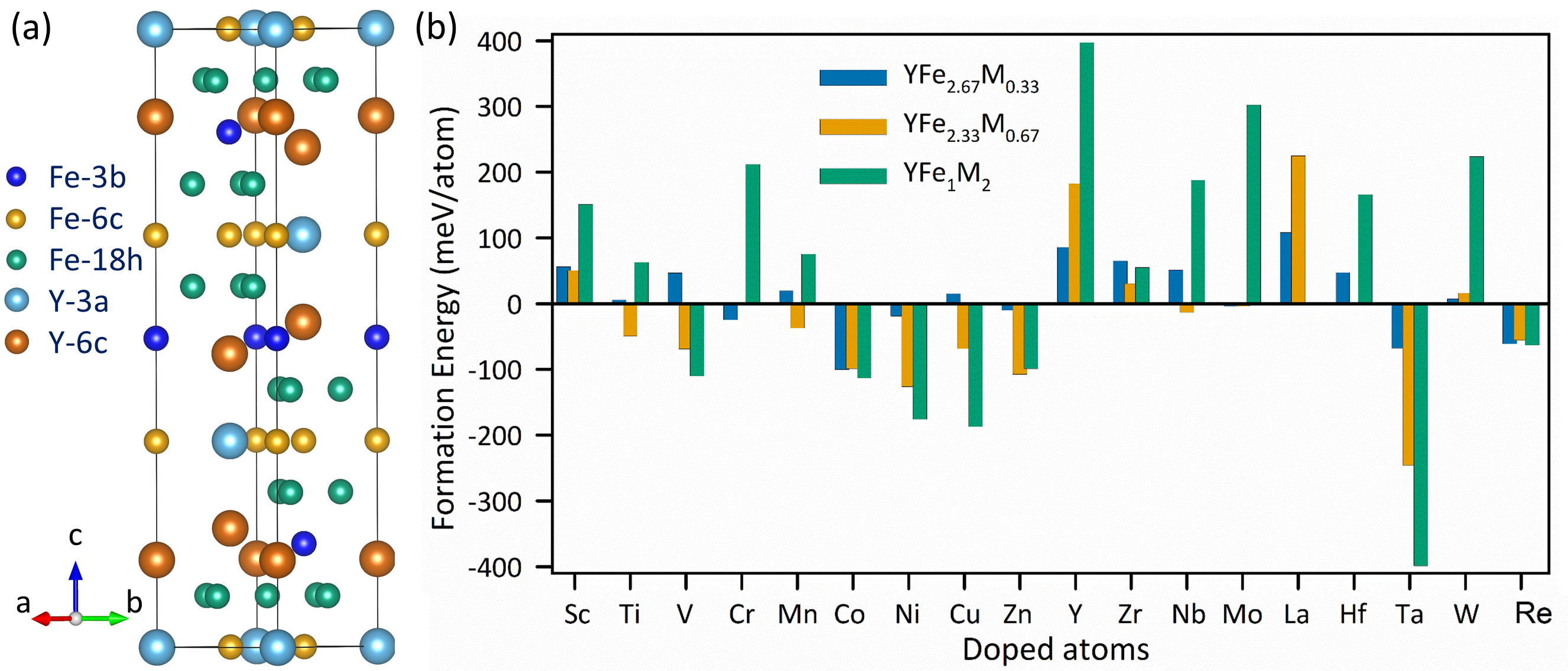}
	\caption{(a) Crystal structure of YFe$_3$ (space group $R\bar{3}m$, No. 166), showing the two inequivalent Y sites (Y-3a, Y-6c) and three Fe sites (Fe-3b, Fe-6c, Fe-18h). (b) Formation energies for elemental substitution at the three inequivalent Fe-sites. Formation energies are calculated from first-principles DFT using GGA exchange-correlation functional, as implemented in VASP, following full relaxation of both lattice parameters and ionic positions.}
	\label{fig:Fig1}
\end{figure*}
%TC:endignore
%%%%%%%%%%%%%%%%%%%%%%%%%%%%%%%%%%%%

\subsection{Micromagnetic simulations}
\label{mm}

The computation of the static hysteresis properties was performed using the Python package \texttt{mammos-mumag} from the MaMMoS software suite~\cite{fangohrMAgneticMultiscaleMOdelling2026, mammosProjectUrl, MaMMoSprojectSupplement2026dopedFe3Ymicromagnetics2026}. \texttt{mammos-mumag} is a finite-element energy minimization tool; it allows simulating hysteresis loops of magnetic materials with multiple grains, enabling the evaluation of extrinsic magnetic properties, including remanence, coercive field, and maximum energy product.

The physical quantity computed with micromagnetic simulations is the unit vector $\textbf{m}$ parallel to the magnetization. The distribution of $\textbf{m}$ is found by minimizing Gibbs's free energy. The magnetization configuration is determined by minimizing the Gibbs free-energy functional, which includes the exchange, magnetocrystalline anisotropy MCA, Zeeman, and magnetostatic energy contributions. The magnetic system is discretized using a finite-element mesh, in which different regions can be assigned distinct intrinsic magnetic parameters, including $M_{\mathrm{s}}$, $K_{\mathrm{u}}$, the exchange stiffness $A$, and the local anisotropy-axis direction. The exchange stiffness $A$ describes the energy associated with spatial variations in magnetization direction and is related to the underlying atomistic exchange interactions $J_{ij}$ in the long-wavelength limit. This approach allows different magnetic regions, including individual grains, grain boundaries, non-magnetic regions, and surrounding computational shells, to be incorporated into the same simulation.
%The energy functional includes the Zeeman, exchange, MCA, and magnetostatic contributions. Gibbs's free energy depends on the local intrinsic properties. The material is discretized using a finite-element mesh, and different mesh regions can be assigned different intrinsic magnetic parameters, such as $M_s$, $K_u$, $A$ and anisotropy-axis directions. This enables homogeneous, multigrain, grain-boundary, non-magnetic, and shell regions to be represented within a single simulation. 
Discretization of Gibbs's free energy translates the problem into an algebraic minimization problem. Its minimization is performed with a preconditioned conjugate gradient method \cite{exlPreconditionedNonlinearConjugate2019a} or the LBFGS method \cite{fischbacherNonlinearConjugateGradient2017a}. The subsequent minimization of Gibbs's free energy for decreasing external field provides the distribution of $\textbf{m}$ along the demagnetization curve. The magnetostatic field is obtained from the magnetic scalar potential, $\mathbf{H}=-\nabla \phi$, and open-boundary conditions are treated using a spherical-shell transformation so that the demagnetizing field decays correctly toward infinity \cite{imhoffOriginalSolutionUnbounded1990}.

Demagnetization curves are calculated by successive energy minimization under a decreasing external magnetic field. At each field step, the equilibrium magnetization from the previous step is used as the initial state for the next minimization, allowing the magnetization reversal process to be followed along the hysteresis branch. The implementation uses the \texttt{JAX} library for automatic differentiation and efficient evaluation of the micromagnetic energy and its gradients.

The input for the micromagnetic simulator is the intrinsic magnetic properties and the temperature-dependent spontaneous magnetization $M_\mathrm{s}(T)$ from ab initio and spin-dynamics simulations. The uniaxial MCA constant $K_\mathrm{u}(T)$ and the exchange stiffness constant $A(T)$ are computed for a certain temperature $T$ using the Kuz'min formula \cite{kuzminExchangeStiffnessFerromagnets2020a}. Additionally, previous micromagnetic calculations on nanocrystalline permanent magnets showed that remanence and coercivity are both strongly dependent on grain size, grain shape, and intergrain coupling \cite{schreflRemanenceCoercivityIsotropic1994, kovacsMicromagneticSimulationsCoercivity2017}. Therefore, the use of several grain sizes is necessary to determine whether the intrinsic anisotropy of doped YFe$_3$ is sufficient to stabilize magnetization reversal in realistic microstructures. For the granular microstructure, we assumed a cube consisting of either columnar grains, equiaxed grains, or platelet-shaped grains (see Fig.~\ref{fig:FigMM}) \cite{kovacsMicromagneticSimulationsCoercivity2017}. The micromagnetic simulations are performed for a cubic magnet geometry. The cube's edge size was varied between 40, 80, 160~nm. The magnetocrystalline anisotropy axes were randomly distributed within a cone with a half-opening angle of 20 degrees. A vacuum was assumed for the intergranular space. The mesh size was 1.5 nm. Simulations were run for the doped systems: YFe$_{2.33}$V$_{0.67}$, YFe$_{2.33}$Nb$_{0.67}$, and YFe$_{2.33}$Ta$_{0.67}$ at 300~K and 450~K. The external field varied from $2H_\mathrm{limit}$ to $-2H_\mathrm{limit}$, with $H_{\mathrm{limit}} = 2K_{\mathrm{u}} / (\mu_{0} M_{\mathrm{s}})$.

The simulated hysteresis loops were post-processed using \texttt{mammos-analysis} Python package \cite{mammosProjectUrl, fangohrMAgneticMultiscaleMOdelling2026} to extract the extrinsic magnetic properties. From each calculated demagnetization curve, the remanence $M_\mathrm{r}$ was obtained from the interpolated magnetization at zero applied field, while the coercive field $H_\mathrm{c}$ was determined from the field at which the magnetization changes sign during the reverse-field branch. For computing the maximum energy product, the magnetization as a function of the external field, $M(H_{\mathrm{ext}})$, is corrected for the demagnetizing field to obtain $M(H)$, where $H$ is the internal magnetic field. Using the macroscopic demagnetizing factor $N=1/3$ for a cube, the internal field is given by
\begin{equation}
    H = H_{\mathrm{ext}} - N M(H_{\mathrm{ext}}),
\end{equation}
and the corresponding magnetic flux density is
\begin{equation}
    B(H) = \mu_0 \left[ H + M(H) \right].
\end{equation}
The $(BH)_{\max}$ value is then calculated as the maximum of all $-BH$ values in the second quadrant of the $B(H)$ curve ($H<0$, $B>0$). This analysis provides the set of extrinsic magnetic quantities, remanent magnetization $M_{\mathrm{r}}$, coercive field $H_{\mathrm{c}}$, and maximum energy product $(BH)_{\max}$.

\subsection{Experimental details}
\label{exp-details}

Y-Fe-based compounds were synthesized by arc melting high-purity (99.9 \% or higher) elemental yttrium, iron, and the selected dopant elements V and Ta. The mixtures were melted in an arc furnace under argon atmosphere, with each sample re-melted at least four times and flipped between melts to ensure chemical homogeneity. After solidification, the ingots were sealed in evacuated quartz ampoules and subjected to heat treatment in a tube furnace when required. 

Crystal structures were examined by X-ray diffraction (XRD) using a PANalytical DY2183 diffractometer equipped with Co K$\alpha$ radiation ($\lambda$ = 1.789 Å). Diffraction patterns were collected over a 2$\theta$ range of 20–120° with a step size of 0.05°. Phase identification was carried out using HighScore software with access to standard reference patterns from the ICDD database. 

Microstructural analysis was performed using a ThermoFisher Scientific Apreo 2S HiVac scanning electron microscope (SEM). Backscattered electron (BSE) images were recorded using an annular BSE detector. Chemical compositions of individual phases were determined by energy-dispersive X-ray spectroscopy (EDS) using an Oxford Instruments Ultim Max 100 silicon drift detector with a 100 mm$^2$ sensor area. Crystallographic orientation analysis was carried out by electron backscatter diffraction (EBSD) in the SEM using an Oxford Instruments C-Nano+ detector operating in resolution mode with a pattern resolution of 1244 × 1024 pixels. The EBSD data were processed using the MTEX toolbox \cite{Bachmann2010} to generate orientation maps and evaluate the angular relationship between the crystallographic c-axis and the sample normal.

Magnetic domain structures were investigated by magneto-optical Kerr microscopy (MOKE). The Kerr microscope was based on a Carl Zeiss Axio Scope microscope equipped with a wide-field Kerr imaging system and a multi-LED illumination source. LEDs were coupled to the microscope through optical fibers arranged in a cross-like configuration, permitting electronic control of the light incidence direction. Depending on the relative orientation between the incidence plane and the magnetization direction, the setup allows polar, longitudinal, and transverse Kerr contrast, which are sensitive to out-of-plane and in-plane magnetization direction. By sequentially pulsing opposite LEDs and subtracting the corresponding images, pure in-plane magnetic domain contrast was obtained while suppressing any polar contribution.

\section{Results and Discussion} 
\label{sec:results} 
\subsection{Dopant Stability and Site Preference}

    YFe$_3$ crystallizes in the rhombohedral PuNi$_3$-type structure with space group $R\bar{3}m$. As shown in Fig.~\ref{fig:Fig1}(a), the crystal structure contains three inequivalent Fe sites (3$b$, 6$c$, and 18$h$) and two inequivalent Y sites ($3a$ and $6c$), providing multiple substitutional environments with distinct local coordination. 
    
    To evaluate the energetics of transition-metal substitution, formation energies $\Delta E_{\mathrm{f}}$ were calculated from fully relaxed DFT structures according to
    \begin{equation}
    \Delta E_{\mathrm{f}}
    =
    E_{\mathrm{tot}}
    -
    \sum_{\alpha} N_{\alpha} E_{\alpha},
    \label{eq:formation_energy}
    \end{equation}
    
    where $E_{\mathrm{tot}}$ is the total energy of the fully relaxed substituted structure, and $N_{\alpha}$ and $E_{\alpha}$ denote the number of atoms and the energy per atom of elemental species $\alpha$, respectively. The resulting formation energies for substitution at the three inequivalent Fe sites are shown in Fig.~\ref{fig:Fig1}(b). Several transition-metal dopants exhibit negative formation energies relative to the elemental reference states, indicating that the corresponding substituted structures are energetically favorable with respect to the constituent elements. The variation in formation energies across the different dopants demonstrates that the substitution energetics are strongly element dependent and cannot be attributed solely to atomic-size effects. This behavior reflects the interplay between local structural relaxation and the electronic structure of the substituting atom, particularly its $3d$-electron occupation. Similar element-dependent trends have been reported for transition-metal substitution in related Fe-based magnetic systems~\cite{Vishina31122025,Vishina02012023}.
 %%%%%%%%%%%%%%%%%%%%%%%%%%%%%%%%%%%%
\begin{figure}[t!]
    \centering
    \includegraphics[width=0.9\columnwidth]{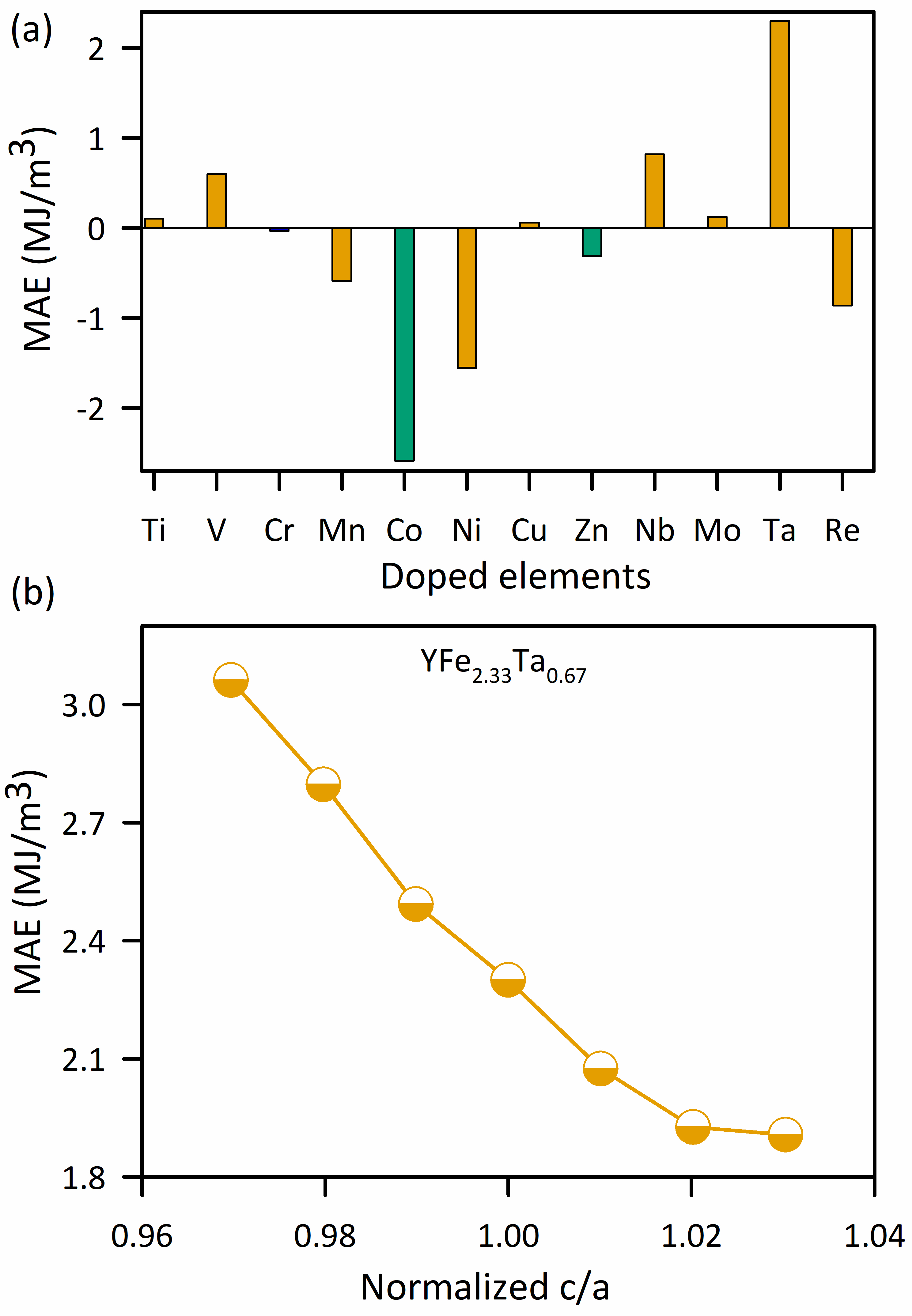}
    \caption{(a) Magnetocrystalline anisotropy energy (MAE) of transition-metal-substituted YFe$_3$ with the dopant occupying the energetically preferred Fe site. The colors follow the same convention as in Fig.~\ref{fig:Fig1}, identifying the corresponding crystallographic sites. Positive (negative) MAE values correspond to uniaxial easy-axis (easy-plane) anisotropy. (b) Dependence of the MAE of YFe$_{2.33}$Ta$_{0.67}$ on the normalized axial ratio, $(c/a)/(c/a)_0$, where $(c/a)_0$ denotes the equilibrium axial ratio. All MAE calculations were performed using the FP-LMTO method implemented in RSPt.}
    \label{fig:mae-all}
\end{figure}
%%%%%%%%%%%%%%%%%%%%%%%%%%%%%%%%%
    To identify the preferred substitutional site, the total energies of configurations containing a single dopant atom at each inequivalent Fe site were compared, as detailed in Supplementary Section~I.B and summarized in Table~S1. Most dopants preferentially occupy the Fe-6$c$ site, whereas Cr favors the Fe-3$b$ site and Co and Zn preferentially substitute at the Fe-18$h$ site. These lowest-energy configurations were subsequently used for the investigation of the magnetic properties. 
    
    For comparison, transition-metal substitution at the Y sublattice was also examined, with the corresponding formation energies presented in Supplementary Section~I.C and Fig.~S2. Comparison reveals a dopant-dependent site preference, with Fe-site substitution being energetically favorable for several of the investigated dopants. These energetic trends motivate the subsequent focus on Fe-site substitution. Because Fe forms the magnetic sublattices of YFe$_3$, substitution at these sites also provides a direct route to modifying the local electronic structure and magnetic interactions that govern the magnetocrystalline anisotropy.
    
\subsection{Dopant-Induced           Magnetocrystalline Anisotropy} 
    The fully relaxed structures obtained from VASP were subsequently employed in MAE calculations performed with the full-potential RSPt code according to Eq.~(\ref{eq:mae}). For each dopant, the energetically preferred Fe substitution site was considered: Fe-6$c$ for most dopants, Fe-3$b$ for Cr, and Fe-18$h$ for Co and Zn. The resulting MAE values are presented in Fig.~\ref{fig:mae-all}(a). Positive MAE values indicate a uniaxial easy axis along the crystallographic $c$ direction, whereas negative values correspond to easy-plane anisotropy. Pristine YFe$_3$ exhibits pronounced easy-plane anisotropy, with an MAE of $-0.90$~MJ\,m$^{-3}$, in good agreement with experimental measurements \cite{bolyachkin2017}.
%%%%%%%%%%%%%%%%%%%%%%%%%%%%%%%%%%%%
\begin{figure}[t!]
    \centering
    \includegraphics[width=0.9\columnwidth]{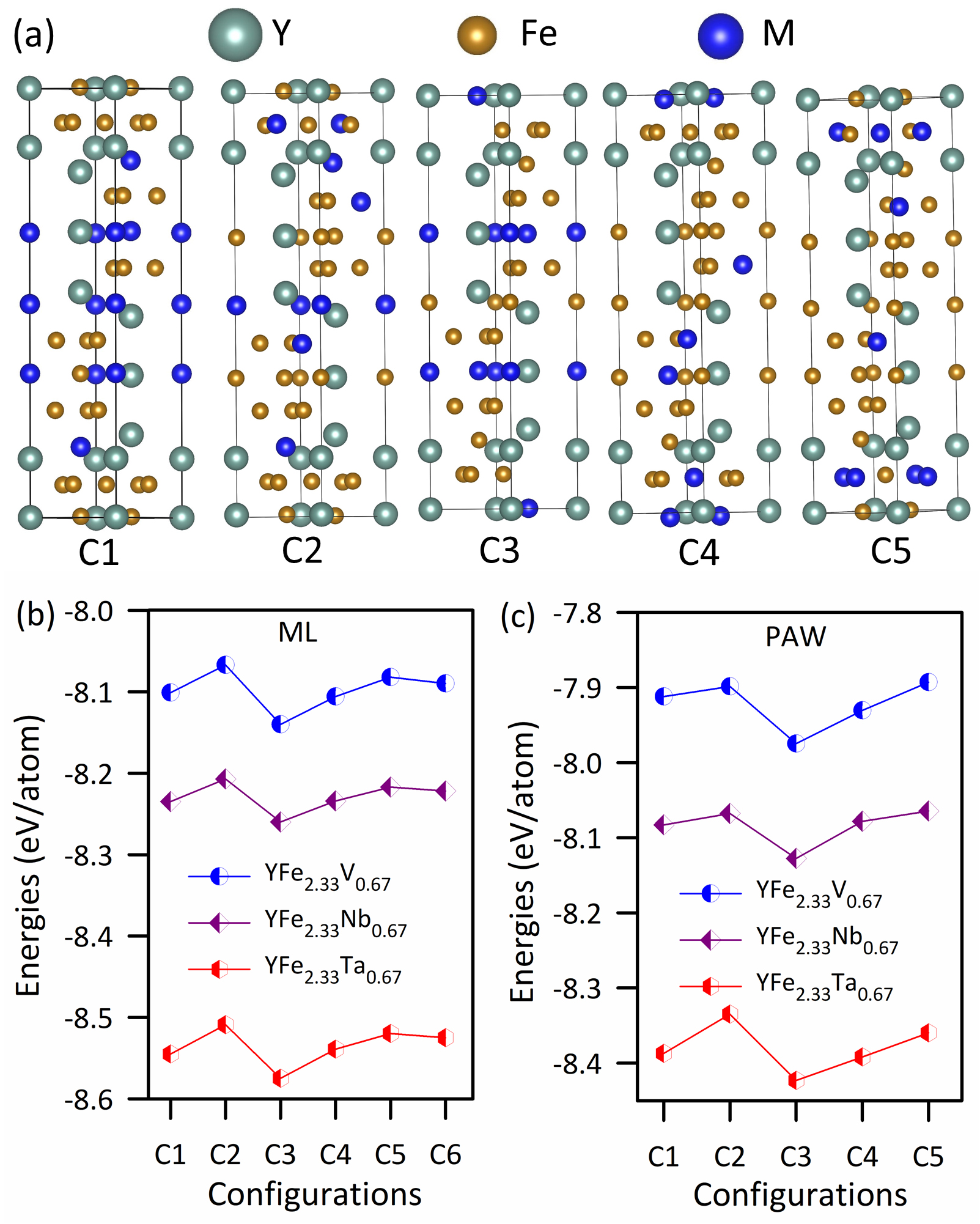}
    \caption{Relative stability of different Fe-sublattice substitution configurations in YFe$_{2.33}$M$_{0.67}$ ($M$ = V, Nb, Ta). (a) Considered ordered and disordered substitutional configurations. C3 denotes the ordered structure, whereas C1, C2, C4, and C5 represent chemically disordered configurations with dopant atoms occupying different combinations of Fe sublattices within the same simulation cell. C6 corresponds to a chemically disordered configuration generated in a $2\times2\times2$ supercell with no constraints on dopant occupation of the Fe sublattices. All structures were generated using the special quasi-random structure (SQS) approach and contain the same dopant concentration. The formation energies were calculated using (b) machine-learned interatomic potentials (MLIPs) in MatterSim and (c) PAW pseudopotentials in VASP.}
    \label{fig:mlip-paw}
\end{figure}
%%%%%%%%%%%%%%%%%%%%%%%%%%%%%%%%%  
    Substitution in one of the Fe-sublattices produces a broad range of anisotropy energies, as shown in Fig.~\ref{fig:mae-all}(a). Several 3$d$ (Ti, V, and Cu) and 4$d$/5$d$ (Nb, Mo, and Ta) dopants reverse the sign of the MAE, driving an easy-plane-to-easy-axis transition. In contrast, Cr, Mn, Co, Ni, Zn, and Re retain negative MAE values and therefore preserve the intrinsic easy-plane magnetic ground state of YFe$_3$. Among all dopants considered, V, Nb, and Ta produce the most pronounced anisotropy enhancement, with Ta yielding the largest MAE of $2.30$~MJ\,m$^{-3}$ and the strongest stabilization of the uniaxial magnetic state.
    
    The observed variation of MAE at the Fe-6c site highlights the strong dependence of the MAE on the electronic structure near the Fermi level, where SOC-induced coupling between occupied and unoccupied states determines the anisotropy energy. Although the strength of spin--orbit coupling generally increases from the 3$d$ to the 5$d$ transition-metal series ($\lambda_{3d}<\lambda_{4d}<\lambda_{5d}$), no simple correlation exists between SOC strength and either the sign or magnitude of the MAE. Nevertheless, several 4$d$ and 5$d$ dopants, particularly Nb and Ta, produce substantial anisotropy enhancements. A notable exception is Re, which retains a negative MAE of $-0.86$~MJ\,m$^{-3}$ despite its strong SOC. These results indicate that the uniaxial anisotropy is determined not only by the strength of SOC but also by the details of the electronic structure near the Fermi level. The microscopic origin of the anisotropy enhancement is discussed in detail in Sec.~\ref{subsec:origin-of-mae}.
     
    For comparison, the MAE values calculated for Y-site substitution are presented in Supplementary Section~I.D and Fig.~S4. Unlike Fe-site substitution, all Y-site-substituted systems retain easy-plane anisotropy. This finding demonstrates that the emergence of uniaxial anisotropy in YFe$_3$ requires direct modification of the magnetic Fe sublattices. 

    Having established that Fe-site substitution modifies the anisotropy of YFe$_3$, we next examine the influence of lattice distortion on the MAE by varying the $c/a$ ratio at constant unit-cell volume. Fig.~2(b) shows that, for YFe$_{2.33}$Ta$_{0.67}$, the MAE increases markedly with decreasing $c/a$, reaching a maximum of $3.00~\mathrm{MJ\,m^{-3}}$ within the investigated range. This pronounced dependence of the MAE on the axial ratio reveals strong magnetoelastic coupling, whereby changes in the crystal-field environment modify the SOC-induced anisotropy. Consequently, lattice distortion enhances the stability of the uniaxial magnetic state. These results demonstrate that strain engineering, in combination with chemical substitution, provides an effective pathway for optimizing the magnetic anisotropy of YFe$_3$-based compounds. The pronounced sensitivity of the MAE to the axial ratio further suggests that epitaxial strain in thin-film geometries may offer an additional route for enhancing the uniaxial anisotropy beyond that of the equilibrium bulk phase.
    
    Given their favorable uniaxial anisotropy (Fig.~\ref{fig:mae-all}a), YFe$_{2.33}$M$_{0.67}$ ($M$ = V, Nb, and Ta) compounds were selected for further study. The dopant atoms were initially placed at the energetically preferred Fe-6$c$ site. To assess the effect of chemical disorder, multiple substitutional configurations were subsequently constructed for each composition (Fig.~\ref{fig:mlip-paw}). Configuration C$_3$ corresponds to the ordered structure, in which all six dopant atoms occupy the Fe-6$c$ sublattice. Configurations C$_1$, C$_2$, C$_4$, and C$_5$ represent partially disordered arrangements, where the six dopant atoms are distributed among different Fe sublattices: Fe-3$b$, Fe-6$c$, and Fe-18$h$. Specifically, C$_1$ and C$_2$ involve the replacement of all three Fe atoms at the Fe-3$b$ site together with three Fe atoms at the Fe-6$c$ and Fe-18$h$ sites, respectively; C$_4$ corresponds to three dopants on the Fe-6$c$ site and three on the Fe-18$h$ site; and C$_5$ corresponds to six dopants occupying Fe-18$h$ positions. C$_6$ represents a fully disordered SQS configuration generated in a $2\times2\times2$ supercell with no restrictions on dopant occupation of the Fe sublattices. All configurations contain the same dopant concentration and differ only in the distribution of dopant atoms among the available Fe sublattices. The relative formation energies of these configurations were first evaluated using the MLIP, enabling efficient screening of the configurational space, and subsequently validated using higher-fidelity DFT calculations with PAW pseudopotentials as implemented in VASP. Figs.~\ref{fig:mlip-paw}b and c compare the relative formation energies obtained from the two approaches. The close agreement between MLIP and DFT, particularly in the relative energies and overall energetic trends, demonstrates the reliability of the MLIP for describing the configurational energetics. In both MLIP and DFT calculations, the ordered C3 configuration exhibits the lowest formation energy, indicating a preference for chemical ordering. This finding is consistent with the single-dopant substitution analysis (Supplementary Table~S1) and confirms that the Fe-6$c$ site remains the preferred substitutional site even in the presence of chemical disorder. Since Fe-6$c$ substitution also yields enhanced uniaxial anisotropy (Table~\ref{tab:mag_properties}), these results establish a direct correlation between site preference and magnetic anisotropy.
 
\subsection{Microscopic Origin of Magnetocrystalline Anisotropy} 
\label{subsec:origin-of-mae} 
\begin{figure}[t!]
	\centering
	\includegraphics[width=0.8\columnwidth]{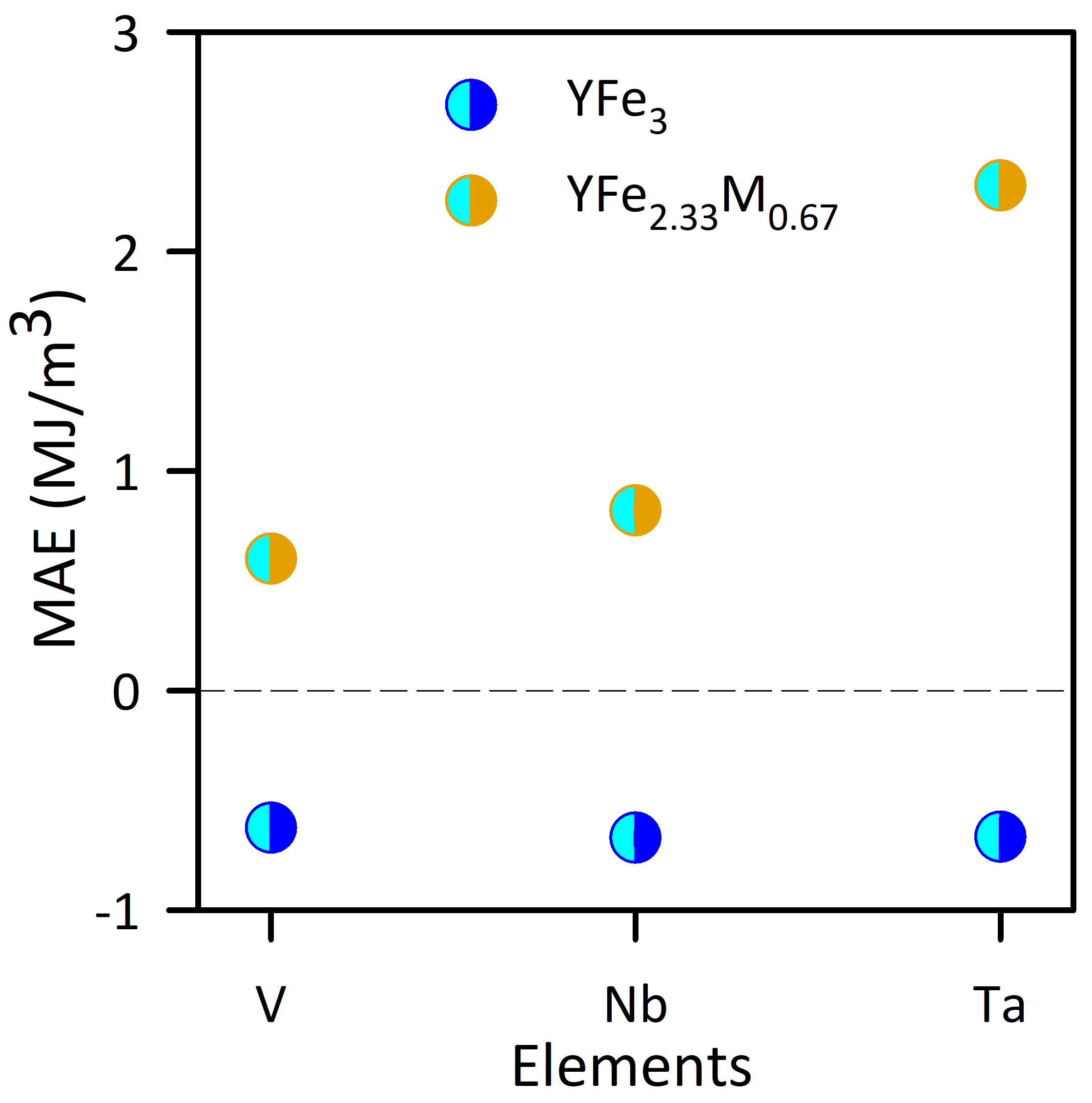}
	\caption{Comparison of the MAE of transition-metal-doped YFe$_3$ and the corresponding undoped YFe$_3$ structures constrained to the relaxed geometries of the doped systems. The persistence of easy-plane anisotropy in the constrained undoped structures indicates that structural relaxation alone is insufficient to induce the easy-axis transition, which instead arises from dopant-induced electronic-structure modifications.}
	\label{fig:Fig3}
\end{figure}

    To elucidate the microscopic origin of the enhanced magnetocrystalline anisotropy in the doped systems, we compare the magnetic anisotropy of YFe$_{2.33}$M$_{0.67}$ ($M$ = V, Nb, and Ta) with that of the corresponding undoped YFe$_3$ structures. To separate the effects of structural relaxation and electronic modification, the MAE was calculated for both doped and undoped systems using the same relaxed atomic geometries obtained for the doped configurations. Fig.~\ref{fig:Fig3} compares the resulting MAE values. Since the doped and undoped systems share identical lattice parameters and atomic geometries, the undoped configurations constrained to the doped structures retain the easy-plane anisotropy characteristic of pristine YFe$_3$. This demonstrates that structural relaxation alone is insufficient to alter the anisotropy direction and affects only the magnitude of the MAE to a limited extent. In contrast, chemical substitution substantially enhances the MAE and drives the transition from easy-plane to uniaxial anisotropy. Therefore, the observed anisotropy enhancement originates primarily from dopant-induced modifications of the electronic structure near the Fermi level, which alter the relevant SOC matrix elements.

%%%%%%%%%%%%%%%%%%%%%%%%%%%%%%%%%%%
%%%%%%%%%%%%%%%%%%%%%%%%%%%%%%%%%%%
\begin{table}[t]
\centering
\caption{Calculated magnetic properties of YFe$_{2.33}$M$_{0.67}$ ($M$ = V, Nb, and Ta). The saturation magnetization ($\mu_0M_\mathrm{s}$), magnetocrystalline anisotropy energy ($K_u$) obtained using the force theorem (FT) and total-energy difference (TE) methods, and Curie temperature ($T_\mathrm{C}$) are listed together with the corresponding values for pristine YFe$_3$.}
\label{tab:mag_properties}
\begin{tabular}{lcccc}
\toprule
\textbf{Compound} 
& \textbf{$\mu_0M_\mathrm{s}$ (T)} 
& \multicolumn{2}{c}{\textbf{$K_u$ (MJ/m$^3$)}} 
& \textbf{$T_\mathrm{C}$ (K)} \\
\cmidrule(lr){3-4}
& 
& \textbf{FT} 
& \textbf{TE} 
& \\
\midrule
YFe$_3$ & 1.08 & $-0.90$ & $-0.91$ & 500 \\
YFe$_{2.33}$V$_{0.67}$ & 0.63 & 0.60 & 0.61 & 555 \\
YFe$_{2.33}$Nb$_{0.67}$ & 0.66 & 0.82 & 0.82 & 490 \\
YFe$_{2.33}$Ta$_{0.67}$ & 0.65 & 2.30 & 2.31 & 535 \\
\bottomrule
\end{tabular}
\end{table}

\begin{figure*}[t!]
	\centering
	\includegraphics[width=0.9\textwidth]{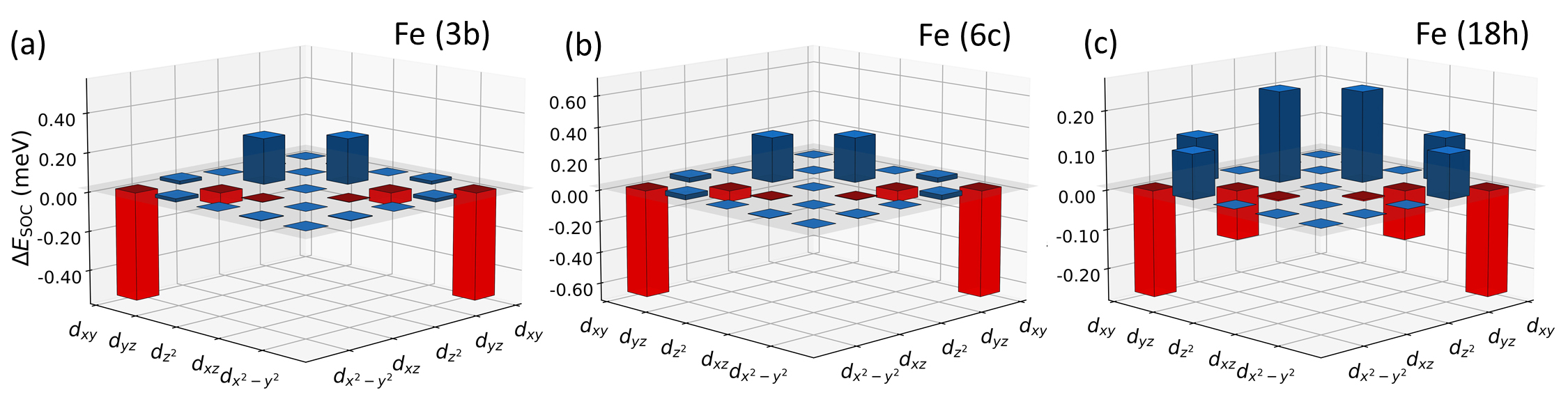}
	\caption{Orbital-resolved SOC-energy anisotropy contributions of the (a) Fe-3$b$, (b) Fe-6$c$, and (c) Fe-18$h$ sites derived from the corresponding spin--orbit coupling energy differences ($\Delta E_{\mathrm{SOC}}$). Results were obtained from GGA-DFT calculations in VASP after full relaxation of lattice parameters and atomic positions.}
	\label{fig:mae_pristin}
\end{figure*}
%%%%%%%%%%%%%%%%%%%%%%%%%%%%%%%%%%%

To gain microscopic insight into the origin of the MCA in pristine and doped YFe$_3$, the SOC contributions were analyzed by resolving the MAE into contributions from individual transition-metal $d$ orbitals. Within the framework of second-order perturbation theory, the SOC-induced contribution to the MAE can be expressed as 

\begin{equation}
E_{\mathrm{MAE}}
= \xi^{2} \sum_{o\alpha,u\beta}
(2\delta_{\alpha\beta}-1)
\frac{
\left|\langle o\beta | L_z | u\alpha \rangle \right|^{2}
-
\left|\langle o\beta | L_x | u\alpha \rangle \right|^{2}
}{
E_{u\alpha} - E_{o\beta}
},
\label{eq:MAE_general}
\end{equation}

    where $\xi$ denotes the SOC strength, $\delta_{\alpha\beta}$ is the Kronecker delta, and $E_{o\beta}$ ($E_{u\alpha}$) represents the eigenenergy of the occupied (unoccupied) $d$-orbital states ($d_{xy}$, $d_{yz}$, $d_{z^{2}}$, $d_{xz}$, and $d_{x^{2}-y^{2}}$) with spins $\beta$ ($\alpha$). The operators $L_x$ and $L_z$ correspond to the $x$ and $z$ components of the orbital angular momentum, respectively (See Supplementary Section~I.E for more details).

    Electronic states located near the Fermi level ($E_\mathrm{F}$) make the largest contribution to the MAE because the energy separation between occupied and unoccupied states appears in the denominator of Eq.~(\ref{eq:MAE_general}). Consequently, smaller energy differences enhance the anisotropy contribution, while the angular-momentum matrix elements in the numerator determine whether the resulting $K_{\mathrm{u}}$ favors easy-plane or uniaxial anisotropy.

    In pristine YFe$_3$, the orbital-resolved density of states (Supplementary Figs.~S5 and S6) shows that the majority-spin $d$ states of Fe (Fe-3$b$, Fe-6$c$, and Fe-18$h$) and Y (Y-3$a$ and Y-6$c$) are nearly fully occupied near the Fermi level, whereas the minority-spin $d$ states remain partially occupied. Consequently, the MAE is governed primarily by SOC-induced transitions between occupied and unoccupied $d$ states in the vicinity of the Fermi level, with the dominant contributions originating from the minority-spin channel owing to the greater availability of low-energy electronic excitations.

    For the Fe-3$b$ and Fe-6$c$ sites (Figs.~\ref{fig:mae_pristin}a,b), the dominant negative contribution to the MAE originates from the antiparallel-spin coupling associated with the matrix element $\langle d_{x^{2}-y^{2}}^{\uparrow} | L_z | d_{xy}^{\downarrow} \rangle$. Additional, but smaller, negative contributions arise from the matrix element $\langle d_{xz}^{\uparrow} | L_z | d_{yz}^{\downarrow} \rangle$. As a result, the total MAE contributions from the Fe-3$b$ and Fe-6$c$ sites amount to $-0.62$ and $-0.67$~meV/atom, respectively, making them the primary origin of the easy-plane anisotropy in pristine YFe$_3$. In contrast, the Fe-18$h$ site (Fig.~\ref{fig:mae_pristin}c) exhibits similar negative contributions from the same orbital couplings. However, these are partially compensated by positive contributions associated with the matrix elements $\langle d_{z^{2}}^{\uparrow} | L_x | d_{yz}^{\downarrow} \rangle$, $\langle d_{x^{2}-y^{2}}^{\uparrow} | L_x | d_{yz}^{\downarrow} \rangle$ and $\langle d_{xy}^{\uparrow} | L_x | d_{xz}^{\downarrow} \rangle$, resulting in a small net positive MAE contribution of approximately $0.10$~meV/atom. In addition to the Fe sites, Y atoms contribute negative MAE values of $-0.19$ and $-0.04$~meV/atom at the Y-3$a$ and Y-6$c$ sites, respectively, thereby favoring easy-plane anisotropy (Supplementary Fig.~S7). Overall, the antiparallel-spin couplings associated with $\langle d_{x^{2}-y^{2}}^{\uparrow} | L_z | d_{xy}^{\downarrow} \rangle$ and $\langle d_{xz}^{\uparrow} | L_z | d_{yz}^{\downarrow} \rangle$ play the dominant role in stabilizing the easy-plane magnetic ground state of pristine YFe$_3$.

\begin{figure*}[t!]
    \centering
    \includegraphics[width=0.9\textwidth]{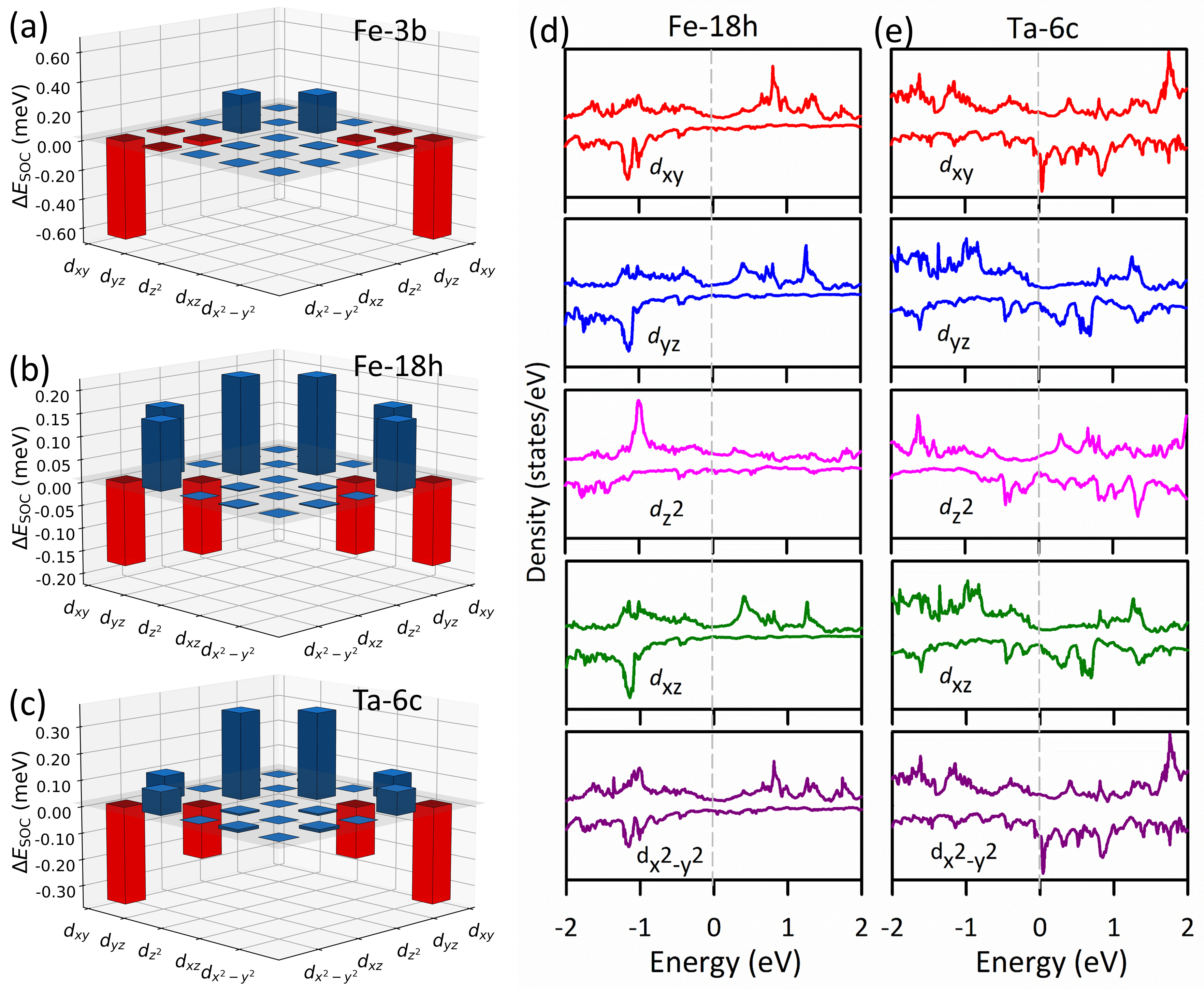}
    \caption{(a-c) Orbital-resolved SOC-energy anisotropy contributions of the (a) Fe-3$b$, (b) Fe~18$h$ and (c) Ta~6$c$ sites in YFe$_{2.33}$Ta$_{0.67}$ system, calculated from the corresponding spin--orbit coupling energy differences ($\Delta E_{\mathrm{SOC}}$). (d-e) $d$-orbital projected partial density of states (PDOS) of YFe$_{2.33}$Ta$_{0.67}$ for (d) Fe~18$h$ site and (e) Ta~6$c$ site from the GGA exchange-correlation functional in VASP, following full relaxation of both lattice parameters and ionic positions.}
    \label{fig:mae-ta}
\end{figure*}

    We focus on YFe$_{2.33}$Ta$_{0.67}$ as a representative case because all three compounds, YFe$_{2.33}$M$_{0.67}$ ($M$ = V, Nb, and Ta), exhibit similar electronic mechanisms underlying the anisotropy enhancement, while Ta substitution yields the largest uniaxial magnetocrystalline anisotropy (Table~\ref{tab:mag_properties}). Corresponding orbital-resolved analyses for YFe$_{2.33}$V$_{0.67}$ and YFe$_{2.33}$Nb$_{0.67}$ are provided in Supplementary Section I.F. The orbital-resolved anisotropy contributions associated with the Ta-6$c$ dopant site are presented in Fig.~\ref{fig:mae-ta}(a--c).
    
    The Fe-3$b$ site retains a negative MAE contribution of $-0.65$~meV/atom, favoring easy-plane anisotropy, similar to pristine YFe$_3$ (Fig.~\ref{fig:mae_pristin}a). This behavior is dominated by the SOC matrix element $\langle d_{x^{2}-y^{2}} | L_z | d_{xy} \rangle$. Smaller positive contributions arise from matrix elements such as $\langle d_{z^{2}}^{\uparrow} | L_z | d_{yz}^{\uparrow} \rangle$. The competition between these positive and negative contributions results in a net easy-plane anisotropy at the Fe-3$b$ site. In contrast, the Ta-6$c$ and Fe-18$h$ sites exhibit positive MAE contributions, favoring uniaxial magnetic anisotropy. Their combined positive contributions outweigh the negative MAE of the Fe-3$b$ site, resulting in the strong overall uniaxial anisotropy of YFe$_{2.33}$Ta$_{0.67}$.
    
    The strong overall uniaxial anisotropy of YFe$_{2.33}$Ta$_{0.67}$ originates primarily from the Ta-6$c$ and Fe-18$h$ sites. Their orbital-resolved MAE contributions and corresponding projected density of states (PDOS) are shown in Fig.~\ref{fig:mae-ta}(a--e). The $d$-projected PDOS of the Fe-18$h$ site (Fig.~\ref{fig:mae-ta}d) shows that the majority-spin states are nearly fully occupied, whereas the minority-spin channel remains partially occupied near the Fermi level. Consequently, SOC-induced transitions involving minority-spin states make significant contributions to the MAE. The orbital-resolved analysis yields a positive MAE contribution of 0.28~meV/atom for the Fe-18$h$ site in YFe$_{2.33}$Ta$_{0.67}$. This positive contribution arises from the combined effect of several competing SOC matrix elements, as shown in Fig.~\ref{fig:mae-ta}b. In particular, the matrix elements $\langle d_{xy} | L_x | d_{xz} \rangle$, $\langle d_{z^{2}} | L_x | d_{yz} \rangle$, and $\langle d_{x^2-y^2} | L_x | d_{yz} \rangle$ provide dominant positive contributions that favor uniaxial anisotropy. These are partially offset by negative contributions associated with couplings such as $\langle d_{x^{2}-y^{2}} | L_z | d_{xy} \rangle$ and $\langle d_{xz} | L_z | d_{yz} \rangle$, resulting in a net positive MAE for the Fe-18$h$ site.

    In contrast, the Ta-6$c$ site contributes 2.9~meV/atom to the anisotropy energy. This contribution originates from SOC-mediated coupling between occupied and unoccupied Ta $d$-states. The matrix elements $\langle d_{xy} | L_x | d_{xz} \rangle$, $\langle d_{z^{2}} | L_x | d_{yz} \rangle$, and $\langle d_{x^{2}-y^{2}} | L_x | d_{yz} \rangle$ represent the dominant positive terms in the second-order perturbation expression, whereas $\langle d_{x^{2}-y^{2}} | L_z | d_{xy} \rangle$ and $\langle d_{xz} | L_z | d_{yz} \rangle$ produce competing negative contributions. The predominance of the positive matrix elements results in a positive anisotropy contribution from the Ta-dopant. Although the pattern matrix-element terms associated with the Fe-18$h$ (Fig.~\ref{fig:mae-ta}b) and Ta-6$c$ (Fig.~\ref{fig:mae-ta}(c)) are qualitatively similar, the Ta-6$c$ site produces a substantially larger positive anisotropy contribution. This behavior originates from the electronic structure near $E_\mathrm{F}$. As shown in Fig.~\ref{fig:mae-ta}e, the occupied and unoccupied Ta $d$ states are located closer to $E_\mathrm{F}$ than the corresponding Fe-18$h$ (Fig.~\ref{fig:mae-ta}d) states. The reduced energy separation between these states decreases the denominator in the second-order perturbation expression (Eq.~(\ref{eq:MAE_general})), thereby amplifying the SOC-induced anisotropy contribution. As a result, the Ta dopant plays a decisive role in stabilizing the uniaxial magnetic state of YFe$_{2.33}$Ta$_{0.67}$.
%%%%%%%%%%%%%%%%%%%%%%%%%%%%%%%%%%%%
%%%%%%%%%%%%%%%%%%%%%%%%%%%%%%%%    
   
    In addition, the Y sites contribute positively to the uniaxial anisotropy, with anisotropy contributions of 0.15~meV/atom and 0.29~meV/atom for the Y-6$c$ and Y-3$a$ sites, respectively, as shown in Supplementary Fig.~S10. Orbital-resolved MAE and $d$-projected PDOS analyses indicate that these contributions originate from distinct SOC-mediated couplings between occupied and unoccupied states near the Fermi level. Although the Y-sites contributions are smaller than the Ta-6$c$ site, they further reinforce the stability of the uniaxial magnetic state in YFe$_{2.33}$Ta$_{0.67}$.
    
    The Fe-6$c$ sites provide the dominant easy-plane contribution in pristine YFe$_3$. Substitution of Fe by Ta at these sites replaces this negative contribution with a strong positive contribution, which, together with the positive contribution from the Fe-18$h$ sites, produces the overall uniaxial anisotropy of YFe$_{2.33}$Ta$_{0.67}$. This reorientation reflects dopant-induced changes in the electronic structure and the balance of SOC-mediated contributions near $E_{\mathrm{F}}$. Similar electronic mechanisms are found in the V- and Nb-substituted compounds (see Supplementary Section I.F).

\subsection{Heisenberg Exchange Interactions and Finite-Temperature Magnetic Properties}
%%%%%%%%%%%%%%%%%%%%%%%%%%%%%%%
\begin{figure*}[t!]
    \centering
    \includegraphics[width=0.85\textwidth]{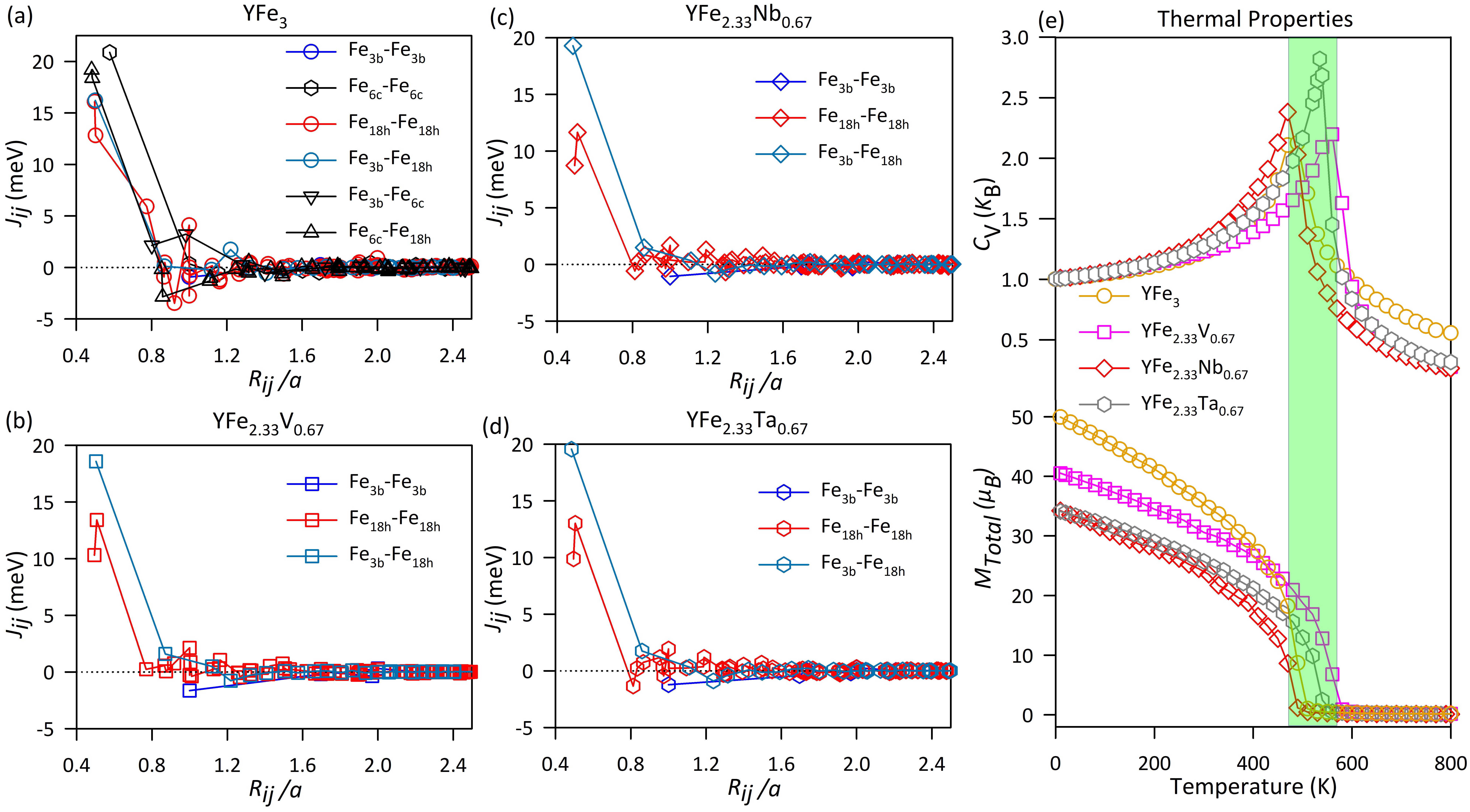}
    \caption{(a--d) Heisenberg exchange parameters ($J_{ij}$) of dominant Fe--Fe exchange interactions as a function of normalized interatomic distance ($R_{ij}/a$) for (a) pristine YFe$_3$ and Fe-site-substituted (b) YFe$_{2.33}$V$_{0.67}$, (c) YFe$_{2.33}$Nb$_{0.67}$, and (d) YFe$_{2.33}$Ta$_{0.67}$. Calculations were performed using the LKAG formalism \cite{LIECHTENSTEIN198765} as implemented in the RSPt code. (e) Temperature-dependent heat capacity ($C_V$) and magnetization of YFe$_3$, YFe$_{2.33}$V$_{0.67}$, YFe$_{2.33}$Nb$_{0.67}$, and YFe$_{2.33}$Ta$_{0.67}$.}
    \label{fig:finite_temp_effect}
\end{figure*}
%%%%%%%%%%%%%%%%%%%%%%%%%%%%%%%%
    To understand the influence of nonmagnetic substitution on the magnetic exchange interactions in YFe$_3$, the calculated Heisenberg exchange parameters ($J_{ij}$) were analyzed as a function of the normalized interatomic distance ($R_{ij}/a$). Fig.~\ref{fig:finite_temp_effect}(a--d) show the dominant exchange interactions between the inequivalent Fe sublattices. The calculated $J_{ij}$ values reveal distinct magnetic behaviors among the inequivalent Fe sublattices, highlighting the site-dependent nature of dopant-induced modifications to the exchange interactions. The rapid decay of $J_{ij}$ with increasing interatomic distance indicates that the magnetic behavior is governed primarily by short-range exchange interactions. For pristine YFe$_3$ (Fig.~\ref{fig:finite_temp_effect}a), the magnetic behavior is dominated by strong ferromagnetic Fe--Fe coupling within the first coordination shell, with the leading exchange parameters exceeding 20~meV. These nearest-neighbor interactions stabilize the ferromagnetic ground state and account for the large spontaneous magnetization and high Curie temperature observed experimentally \cite{bolyachkin2017,givord1971}. Beyond the first coordination shell, the exchange interactions decrease rapidly in magnitude and exhibit only weak oscillations around zero, indicating a minor contribution from longer-range interactions. Upon substitution (Figs.~\ref{fig:finite_temp_effect}(b--d)), the Fe-3$b$--Fe-3$b$ interaction remains nearly unchanged and retains a weak antiferromagnetic character. However, because the number of such pairs is small compared with the dominant ferromagnetic Fe--Fe interactions, their contribution to the overall magnetic behavior is negligible.

    In YFe$_{2.33}$V$_{0.67}$ (Fig.~\ref{fig:finite_temp_effect}b), the Fe-18$h$--Fe-18$h$ and Fe-3$b$--Fe-18$h$ interactions remain strongly ferromagnetic, whereas the Fe--V interactions are weakly antiferromagnetic at short distances and the V--V interactions are negligible. Notably, the nearest-neighbor Fe-18$h$--Fe-18$h$ interaction is reduced compared with pristine YFe$_3$, such that the second-neighbor Fe-18$h$--Fe-18$h$ exchange becomes the dominant interaction within this sublattice. This behavior reflects the modification of the local exchange interactions arising from the altered atomic environment and Fe–Fe distances introduced by V substitution, while preserving the overall ferromagnetic exchange network. Similarly, YFe$_{2.33}$Nb$_{0.67}$ (Fig.~\ref{fig:finite_temp_effect}c) exhibits weak antiferromagnetic Fe--Nb couplings at short range, while the dominant Fe--Fe interactions remain ferromagnetic. As in the V-substituted compound, the nearest-neighbor Fe-18$h$--Fe-18$h$ exchange is weakened relative to the second-neighbor interaction, reflecting a modification of the local exchange pathways around the substituted sites. Nevertheless, the overall exchange balance continues to favor a ferromagnetic ground state. Ta substitution (Fig.~\ref{fig:finite_temp_effect}d) largely preserves the Fe–Fe exchange network while producing the largest MAE.. The Fe--Fe exchange network remains largely unchanged from that of pristine YFe$_3$, although the nearest-neighbor Fe-18$h$--Fe-18$h$ interaction is again somewhat suppressed relative to the second-neighbor coupling. The Fe--Ta interactions are weakly antiferromagnetic and small in magnitude. Despite the presence of Fe 3$d$--Ta 5$d$ hybridization, the dominant ferromagnetic Fe--Fe interactions are preserved. Consequently, Ta minimally perturbs the magnetic exchange while substantially enhancing the spin--orbit-coupling-driven anisotropy, making it particularly effective for improving magnetocrystalline anisotropy without compromising the ferromagnetic ground state.

    To further investigate the finite-temperature magnetic properties, atomistic Monte Carlo simulations were carried out using the classical spin Hamiltonian [Eq.~(\ref{eq:heisenberg})]. The calculated magnetization, $M(T)$, and heat capacity, $C_V(T)$, are presented in Fig.~\ref{fig:finite_temp_effect}(e). The Curie temperature ($T_\mathrm{C}$) was extracted from the maximum of $C_V(T)$, corresponding to the ferromagnetic-to-paramagnetic phase transition. For comparison, the corresponding mean-field Curie temperatures obtained from the calculated exchange interactions are summarized in Supplementary Table~S2. Pristine YFe$_3$ exhibits a Curie temperature of approximately \SI{500}{\kelvin}, in good agreement with experimental reports \cite{givord1971}. This relatively high transition temperature originates from the strong nearest-neighbor ferromagnetic exchange interactions identified in the $J_{ij}$ analysis. Upon transition-metal substitution, the low-temperature magnetization is reduced to varying degrees, reflecting the replacement of magnetic Fe atoms by dopant species. Nevertheless, only moderate changes in thermal magnetic stability are observed. YFe$_{2.33}$V$_{0.67}$ exhibits a slightly higher $T_\mathrm{C}$ of approximately \SI{555}{\kelvin}, whereas Nb substitution reduces $T_\mathrm{C}$ to about \SI{490}{\kelvin}. In contrast, YFe$_{2.33}$Ta$_{0.67}$ retains a high $T_\mathrm{C}$ of \SI{535}{\kelvin} while simultaneously exhibiting strong uniaxial anisotropy. Overall, all doped compounds maintain transition temperatures close to \SI{500}{\kelvin}, indicating that the dominant Fe--Fe exchange interactions remain largely preserved upon transition-metal substitution. Notably, the Ta-substituted system combines enhanced spin--orbit-driven anisotropy with robust high-temperature ferromagnetism, making it a particularly promising candidate for permanent-magnet applications.  

% \begin{table*}[t]
% \centering
% \caption{Summary of micromagnetic extrinsic properties for doped YFe$_3$ systems.}
% \label{tab:extrinsic_summary}
% \begin{tabular}{lccccl}
% \hline
% Dopant & Temperature & Mean $(BH)_\mathrm{max}$ & Mean $\mu_0 H_\mathrm{c}$ & Mean $\mu_0 M_\mathrm{r}$ & Main conclusion \\
%        & (K) & (kJ m$^{-3}$) & (T) & (T) & \\
% \hline
% Ta & 300 & 43.78 & 2.49 & 0.47 & Best coercive stability \\
% V  & 300 & 43.17 & 0.53 & 0.47 & Similar $(BH)_\mathrm{max}$, much lower $H_\matherm{c}$ \\
% Nb & 300 & 33.79 & 0.63 & 0.42 & Lower energy product and moderate $H_\matherm{c}$ \\
% Ta & 450 & 19.13 & 1.12 & 0.31 & Best high-temperature coercivity \\
% V  & 450 & 23.78 & 0.29 & 0.35 & Best high-temperature $(BH)_\mathrm{max}$, but low $H_\matherm{c}$ \\
% Nb & 450 & 7.71  & 0.13 & 0.20 & Weak high-temperature performance \\
% \hline
% \end{tabular}
% \end{table*}

% \begin{table}[t]
% \centering
% \caption{Summary of micromagnetic extrinsic properties for doped YFe$_3$ systems.}
% \label{tab:extrinsic_summary}
% \begin{tabular}{lcccc}
% \hline
% Dopant & Temperature & Mean $(BH)_\mathrm{max}$ & Mean $\mu_0 H_\mathrm{c}$ & Mean $\mu_0 M_\mathrm{r}$ \\
%        & (K) & (kJ m$^{-3}$) & (T) & (T) \\
% \hline
% Ta & 300 & 43.78 & 2.49 & 0.47 \\
% V  & 300 & 43.17 & 0.53 & 0.47 \\
% Nb & 300 & 33.79 & 0.63 & 0.42 \\
% Ta & 450 & 19.13 & 1.12 & 0.31 \\
% V  & 450 & 23.78 & 0.29 & 0.35 \\
% Nb & 450 & 7.71  & 0.13 & 0.20 \\
% \hline
% \end{tabular}
% \end{table}

\subsection{Micromagnetic simulations}
\begin{figure*}[t!]
	\centering
	\includegraphics[width=\textwidth]{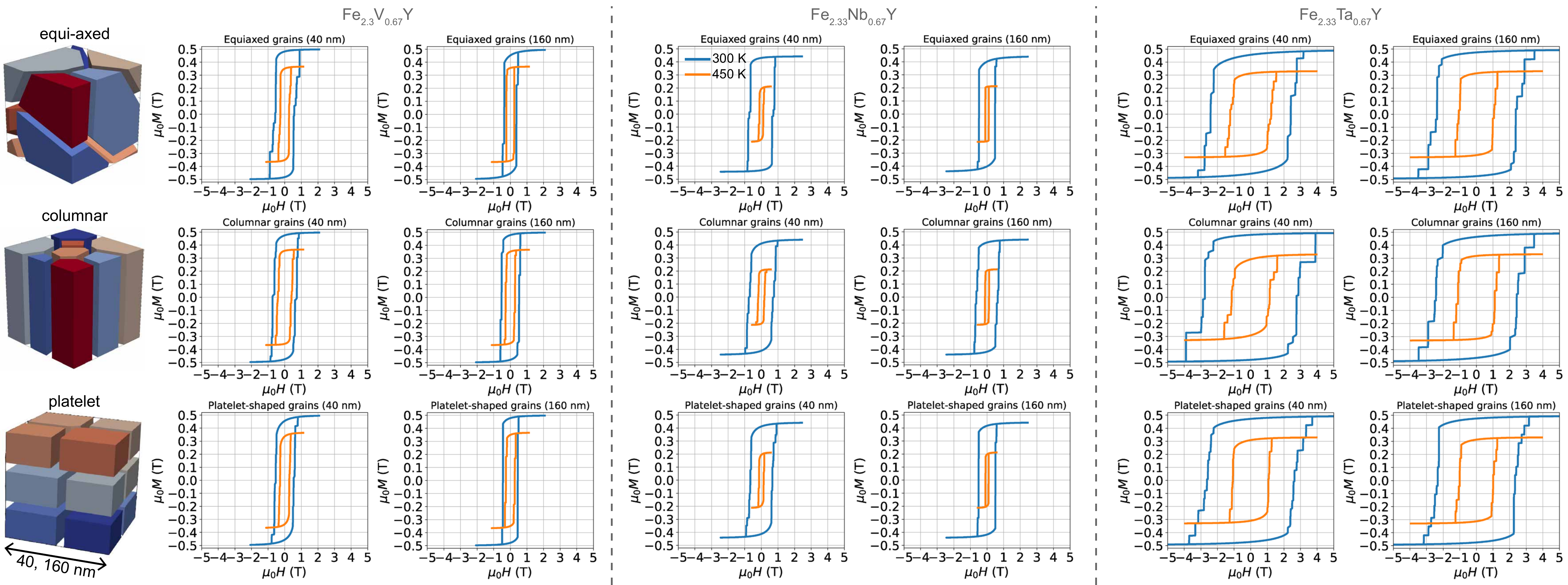}
	\caption{Computed hysteresis loops for YFe$_{2.33}$V$_{0.67}$, YFe$_{2.33}$Nb$_{0.67}$, and YFe$_{2.33}$Ta$_{0.67}$ for different microstructures. From left to right for each system the simulated cube size is increasing from 40~nm to 160~nm. From top to bottom the grain type is varied between equiaxed, columnar, and platelet. The blue and the orange curves refer to 300~K and 450~K, respectively.}
	\label{fig:FigMM}
\end{figure*}

    To assess the impact of the calculated intrinsic magnetic properties on macroscopic permanent-magnet performance, finite-element micromagnetic simulations were performed for representative grain morphologies and temperatures. Fig.~\ref{fig:FigMM} shows that the three doped YFe$_3$ systems have different hysteresis behaviors. The Ta-doped compound has the widest loops and the largest coercive fields over the full set of simulated microstructures. At 300~K, YFe$_{2.33}$Ta$_{0.67}$ reaches $\mu_0H_\mathrm{c}=2.76$~T for 40~nm cube size with columnar grains, while maintaining $\mu_0M_\mathrm{r}=0.48$~T and $(BH)_\mathrm{max}=45.57$~kJ/m$^3$. At 450~K, the remanence and energy product are reduced, but the coercivity remains above 1~T for all simulated Ta-containing microstructures. This indicates that Ta doping gives the best resistance to irreversible demagnetization among the tested dopants.

The V-doped compound gives a different ranking depending on whether $(BH)_\mathrm{max}$ or $H_\mathrm{c}$ is used as the main criterion. At 300~K, YFe$_{2.33}$V$_{0.67}$ reaches a maximum $(BH)_\mathrm{max}=45.63$~kJ/m$^3$, essentially the same as the best Ta-doped case. However, its maximum coercive field is only $\mu_0H_\mathrm{c}=0.73$~T, obtained for 40~nm columnar grains. Thus, $(BH)_\mathrm{max}$ alone would suggest that V- and Ta-doped systems are comparable at room temperature, whereas the coercive field shows that Ta-doped YFe$_3$ is substantially more stable against opposing fields. At 450~K, the V-doped system has the largest calculated energy product among the three dopants, $(BH)_\mathrm{max}=24.98$~kJ/m$^3$ for 40~nm columnar grains, but the corresponding coercive field is only $\mu_0H_\mathrm{c}=0.41$~T. This makes V doping attractive from the energy-density perspective, but less robust for applications where strong demagnetizing fields or high-temperature irreversible losses are expected.

The Nb-doped compound shows the weakest overall hard-magnetic behavior. At 300~K, its best energy product is $36.69$~kJ/m$^3$, and its largest coercive field is $0.74$~T. At 450~K, both remanence and coercivity are strongly reduced. The best energy product is only $8.60$~kJ/m$^3$, and the highest coercive field is $0.19$~T. In several Nb-doped cases at 450~K, $\mu_0H_\mathrm{c}$ becomes comparable to $\mu_0M_\mathrm{r}/2$, indicating that the energy product is close to becoming coercivity limited. This behavior is consistent with the general micromagnetic criterion that the theoretical remanence-limited value of $(BH)_\mathrm{max}$ cannot be reached when the coercive field is too small to sustain a square second-quadrant demagnetization curve \cite{fischbacherMicromagneticsRareearthEfficient2018a, kovacsMicromagneticSimulationsCoercivity2017}.

Calculated from hysteresis loops (see Fig.~\ref{fig:FigMM} and Figs.~S11-S13 in Supplementary materials) values of extrinsic magnetic parameters (see Tables~S3-S5 in Supplementary materials) show that $(BH)_\mathrm{max}$ mostly follows the remanence, whereas $H_\mathrm{c}$ is more sensitive to dopant chemistry, grain shape, and grain size. For most simulated microstructures, the calculated $(BH)_\mathrm{max}$ is close to the ideal remanence-limited estimate
\begin{equation}
    (BH)_\mathrm{max}^{\mathrm{ideal}}
    \approx
    \frac{(\mu_0M_\mathrm{r})^2}{4\mu_0},
\end{equation}
which means that the energy product is primarily limited by the modest remanence of these YFe$_3$-based phases. Nevertheless, the coercive field remains decisive for practical use. This is most clearly seen by comparing Ta- and V-doped YFe$_3$ at 300~K: both reach approximately $45$~kJ/m$^3$, but the best Ta-doped magnet has nearly four times larger coercivity than the best V-doped magnet. Therefore, Ta doping is favored when resistance to demagnetizing fields is required, whereas V doping provides competitive energy density but lower magnetic stability.

The grain-shape dependence further confirms the importance of microstructure. Columnar grains generally provide the highest coercivity, especially for V- and Nb-doped systems, because their elongated shape reduces the demagnetizing field along the easy axis and contributes a favorable shape anisotropy. However, the highest $(BH)_\mathrm{max}$ is not always obtained for the highest-coercivity microstructure, because $(BH)_\mathrm{max}$ also depends strongly on $\mu_0M_\mathrm{r}$ and the shape of the second-quadrant loop. Consequently, optimizing these YFe$_3$-based materials requires a simultaneous balance of remanence, coercivity, and microstructural stability rather than maximizing $(BH)_\mathrm{max}$ alone.

\subsection{Experimental Validation}

To assess the experimental accessibility of the predicted anisotropy reorientation, we investigated V and Ta incorporation into YFe$_3$. Despite yielding the largest calculated uniaxial anisotropy, Ta showed no appreciable incorporation into the YFe$_3$ phase under the synthesis conditions employed. We therefore focus on V-containing alloys, beginning with the nominal composition YFe$_{2.33}$V$_{0.67}$ examined computationally and subsequently exploring lower V concentrations. In samples with measurable V incorporation into YFe$_3$, correlative EBSD and Kerr microscopy were used to relate crystallographic orientation to magnetic domain structure and assess the magnetic anisotropy.

\begin{figure}[htbp] 
    \centering \includegraphics[width=0.95\columnwidth]{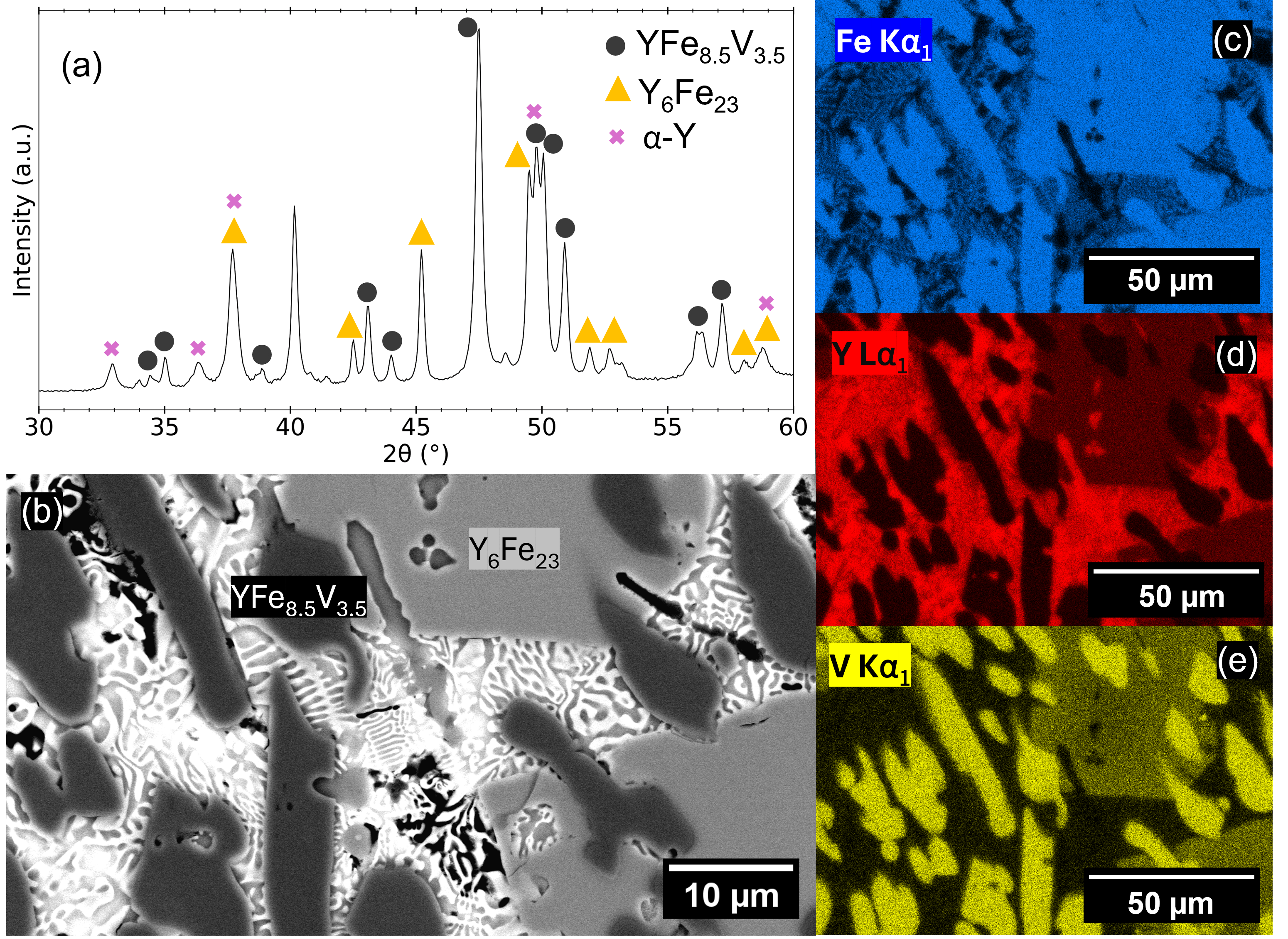} 
    \caption{XRD, SEM, and EDS characterization of the as-arc-melted YFe$_{2.33}$V$_{0.67}$ sample. Panel (a) displays the XRD pattern, highlighting the formation of YFe$_{8.5}$V$_{3.5}$, Y$_6$Fe$_{23}$, and $\alpha$-Y, with no evidence of YFe$_3$. Panel (b) shows a BSE image revealing a multiphase microstructure. The dark regions are identified as YFe$_{8.5}$V$_{3.5}$, whereas the intermediate gray regions correspond to Y$_6$Fe$_{23}$. The brightest regions in the BSE image are associated with eutectic areas consisting of $\alpha$-Y and a Y-Fe phase. Panels (c), (d), and (e) present the corresponding EDS elemental maps for Fe K$\alpha_1$, Y L$\alpha_1$, and V K$\alpha_1$ respectively.} 
    \label{fig:Fig8} 
\end{figure}

Fig.~\ref{fig:Fig8} presents the structural and microstructural characterization of an as-arc-melted sample with nominal composition YFe$_{2.33}$V$_{0.67}$, analyzed by XRD, SEM, and EDS. The XRD pattern shown in Fig.~\ref{fig:Fig8}(a) indicates the formation of a multiphase bulk material, with the main phases identified as YFe$_{8.5}$V$_{3.5}$ and Y$_6$Fe$_{23}$, along with minor contributions from $\alpha$-Y. No diffraction peaks corresponding to YFe$_3$ were observed in the experimental XRD pattern.

Fig.~\ref{fig:Fig8}(b) illustrates the microstructural features acquired in BSE mode of the as-arc-melted V-doped sample. A complex multiphase microstructure is observed, primarily composed of YFe$_{8.5}$V$_{3.5}$  (darker gray) and Y$_6$Fe$_{23}$ (intermediate gray), together with eutectic regions (brighter), consisting of Y-Fe and Y-rich phases. Under the employed synthesis conditions, the YFe$_3$ phase was not detected. However, as evidenced by the EDS elemental maps in panels~(c)--(e) of Fig.~\ref{fig:Fig8}, V exhibits appreciable solubility in the Y-Fe system. Notably, point EDS analysis of the cubic Y$_6$Fe$_{23}$ phase indicates a V solubility of approximately 10 at. \% (data not shown). Nevertheless, under equilibrium solidification conditions following arc melting, phase formation favors the tetragonal ThMn$_{12}$-type (1:12) structure over the rhombohedral PuNi$_3$-type (1:3) structure in the V-doped Y-Fe system. 

The tetragonal 1:12 structure is not stable in the binary Y-Fe phase diagram, as it would normally decompose into a mixture of tetragonal 2:17 and bcc $\alpha$-Fe phases. The formation of the YFe$_{8.5}$V$_{3.5}$ phase observed in this study is consistent with previous reports highlighting the stabilizing effect of alloying elements in rare-earth-Fe systems. For example, Chen \textit{et al.} (2026) \cite{Chen2026} demonstrated that V acts as an effective stabilizer in rare-earth 1:12 (Sm-Fe-based) compounds. Specifically, in Y-Fe-V alloys, Shunkang Pan \textit{et al.} (2012) \cite{Pan2012} reported the formation of YFe$_{12-x}$V$_x$ phase ($1.5 < x < 2.7$) with a tetragonal structure (space group I4/mmm), further supporting the preference for 1:12 phase formation in the presence of V.

More broadly, the stabilization of the 1:12 phase in Y-Fe systems has been attributed to several elements, including Mo, Ti, and V. Isnard \textit{et al.} (2009) \cite{Isnard2009} demonstrated that V can be incorporated into YFe$_{12-x}$V$_x$ for compositions up to x = 3.5, confirming both the solubility of V in the Y-Fe system and its stabilizing effect on the 1:12 phase. Similarly, Hong Sun \textit{et al.} (1993) \cite{Sun1993} discussed the stabilization of the 1:12 phase by Mo and its influence on magnetic properties. Additional studies, such as Bessais (2021) \cite{Review2022}, further support the role of Ti and Mo as key stabilizing elements for this structure. The formation of 1:12 phases has also been observed in related rare-earth systems: Maccari \textit{et al.} (2019) \cite{Maccari2019} successfully produced the 1:12 phase in Nd-Fe-Ti alloys, while Dirba \textit{et al.} (2020) \cite{Dirba2020} reported an investigation in Sm-Fe-Ti system.

From the magnetic property perspective, the 1:12 phase is particularly attractive for permanent magnet applications due to its uniaxial magnetocrystalline anisotropy with a well-defined easy axis \cite{Verhoef1988}. Additionally, Buschow (1991) \cite{Buschow1991} reported a Curie temperature of approximately \SI{550}{\kelvin} for this phase, highlighting its potential for magnetic applications. It should be noted, however, that no magnetic measurements were performed in the present work, as reliable assessment is challenging in multiphase samples. For detailed magnetic characterization of the 1:12 phase, the reader is referred to the cited literature.
\begin{figure*}[t!]
	\centering
	\includegraphics[width=0.9\textwidth]{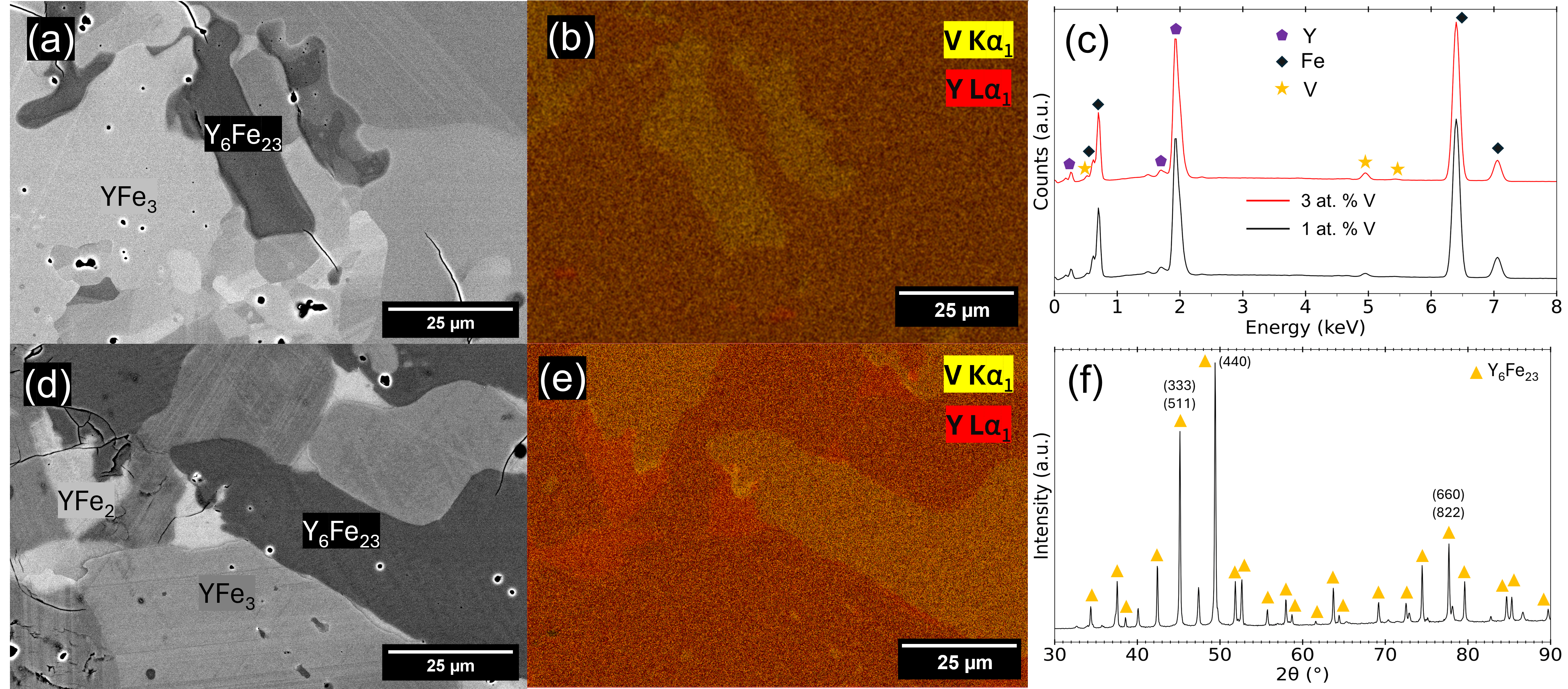}
	\caption{Characterization of Y-Fe samples doped with a few atomic percent V after heat treatment. (a) BSE image of the sample with 1 at. \% V, showing a YFe$_3$ matrix (bright gray) with minor Y$_6$Fe$_{23}$ precipitates (dark gray). (b) Corresponding EDS elemental maps (Y L$\alpha_1$ and V K$\alpha_1$), indicating V solubility in YFe$_3$ phase. (c) Representative EDS spectra confirming up to 2 at. \% V in YFe$_3$. (d) BSE image of the sample with 3 at. \% V, revealing a multiphase microstructure consisting of Y$_6$Fe$_{23}$ (dark gray), YFe$_3$ (intermediate gray), and minor YFe$_2$ (bright gray). (e) Corresponding EDS elemental maps (Y L$\alpha_1$ and V K$\alpha_1$), showing V distribution with higher concentration in Y$_6$Fe$_{23}$. (f) XRD pattern of the sample with 5 at. \% V, identifying Y$_6$Fe$_{23}$ as the predominant phase.}
	\label{fig:Fig9}
\end{figure*}
A new series of samples was prepared to promote the formation and stabilization of V as a dopant in the YFe$_3$ phase. This was achieved by alloying with a few atomic percent of V (1, 3, and 5 at. \%), followed by a two-step heat treatment consisting of homogenisation at \SI{1050}{\celsius} for 5 days, then annealing at \SI{1200}{\celsius} for 2 hours followed by quench on water, aiming to favor the formation and retention of the YFe$_3$ structure. For more investigation regarding heat treatment on binary Y-Fe alloy, the reader is referred to the supplementary material.

For the heat-treated sample with an initial composition of 1 at. \% V, the microstructure (Fig~\ref{fig:Fig9}) reveals a multiphase structure composed of YFe$_3$ grains with minor Y$_6$Fe$_{23}$ precipitates. Panel (a) shows a detailed BSE image, in which the YFe$_3$ phase is visible (brighter gray), while Y$_6$Fe$_{23}$ is observed as darker gray precipitates. BSE intensity variations within the YFe$_3$ grains likely reflect differences in crystallographic orientation, typical of isotropic grains in BSE imaging. Pores are also visible, evident in black, round features in the micrograph.

The corresponding EDS elemental maps for Y L$\alpha_1$ and V K$\alpha_1$ (Fig.~\ref{fig:Fig9}(b)) show that V is distributed throughout the analyzed area, indicating measurable solubility within the YFe$_3$ phase. Point EDS analysis, illustrated by the black spectrum in panel (c), confirms the incorporation of approximately 1 at. \% V into the YFe$_3$ phase and about 2 at. \% V into the secondary Y$_6$Fe$_{23}$ phase (data not shown).

The SEM characterization of the as-heat-treated sample with an initial composition of 3 at. \% V (Fig.~\ref{fig:Fig9}(d)) reveals a multiphase microstructure in BSE mode. The darker gray regions correspond to Y$_6$Fe$_{23}$, the intermediate gray to YFe$_3$, and the minor brighter gray phase is identified as YFe$_2$. The EDS elemental maps of Y L$\alpha_1$ and V K$\alpha_1$ (Fig.~\ref{fig:Fig9}(e)) show that V is distributed throughout the analyzed region, with a higher concentration in the cubic Y$_6$Fe$_{23}$ phase. Point EDS analysis, illustrated by the red spectrum in Fig.~\ref{fig:Fig9}(c), confirms approximately 2 at. \% V incorporation in YFe$_3$, 4 at. \% in Y$_6$Fe$_{23}$, and about 2 at. \% in YFe$_2$ (data for 6:23 and 1:2 phases not shown).

For the as-heat-treated sample with an initial composition of 5 at. \% V, Fig.~\ref{fig:Fig9}(f) presents the corresponding structural characterization. The XRD pattern shows that the cubic Y$_6$Fe$_{23}$ phase is predominant, with the three most intense diffraction peaks indexed and the corresponding crystallographic planes and 2$\theta$ positions indicated. Point EDS analysis reveals a V content of approximately 5.7 at. \% in the Y$_6$Fe$_{23}$ phase. Neither XRD nor SEM-EDS analysis detected the presence of the YFe$_3$ phase in this sample, indicating that 5 at. \% V may exceed the solubility limit for incorporation into YFe$_3$ under the applied synthesis conditions. Under the present synthesis and heat-treatment conditions, V incorporation into YFe$_3$ was observed up to approximately 2~at.\%.

A similarly low V solubility has been reported in the structurally related YNi$_3$ phase. Pan \textit{et al.} \cite{Pan2009} found a maximum V solubility of approximately 2~at.\% in YNi$_3$ at \SI{773}{\kelvin}. Although the Y--Ni--V and Y--Fe--V systems differ chemically and were investigated under different processing conditions, this result provides a useful structural comparison and is consistent with the limited V incorporation observed here in YFe$_3$. In the present Y--Fe--V samples, excess V is instead accommodated by the formation of secondary phases such as Y$_6$Fe$_{23}$ and YFe$_2$.

To relate the experimentally accessible V concentrations to the predicted magnetic anisotropy, we additionally calculated the concentration dependence of the MAE in V-substituted YFe$_3$. The calculated MAE increases systematically with increasing V content and changes sign between approximately 11 and 14~at.\% V, indicating an easy-plane-to-easy-axis transition only at substantially higher V concentrations than those achieved experimentally. The corresponding concentration-dependent MAE values are shown in Supplementary Fig.~S14. Thus, the experimentally observed V incorporation of up to approximately 2~at.\% remains well within the calculated easy-plane regime.

To determine whether V incorporation into the YFe$_3$ phase (up to 2 at. \%) induces a transition from easy-plane to easy-axis magnetocrystalline anisotropy, a correlative EBSD-Kerr microscopy approach was employed. The correlative EBSD--Kerr approach enables direct comparison between crystallographic orientation and local magnetic domain structures. EBSD-based orientation analysis has been widely applied to permanent-magnet systems, including Nd--Fe--B \cite{Bittner2017,Hrkac2014,Woodcock2012,Woodcock2011}, Sm--Co \cite{Yang2021,Yonamine2008}, and Mn--Al-based magnets \cite{Gusenbauer2020,Bittner2017a,Bittner2015}. Correlative crystallographic and magnetic-domain imaging has also been employed in other magnetic materials \cite{Santa-aho2026,Nadoum2020,Batista2014}. In the present work, EBSD was first used to identify YFe$_3$ grains with the crystallographic $c$-axis oriented normal to the sample surface, followed by Kerr microscopy on the same grains.

%This enables a direct correlation between crystallographic orientation and local magnetic domain patterns. The application of EBSD as a microstructural characterization tool is well established across a wide range of material systems. For example, it has been applied to high-energy-product magnets based on the Nd-Fe-B system \cite{Bittner2017, Hrkac2014, Woodcock2012, Woodcock2011} and Sm-Co \cite{Yang2021, Yonamine2008}, as well as rare-earth-free permanent magnets Mn-Al-based system \cite{Gusenbauer2020, Bittner2017a, Bittner2015}. Furthermore, the combination of crystallographic characterization with magnetic domain imaging has been successfully applied in other application context, including electrical steels and cementite, demonstrating the correlation between crystallographic orientation and magnetic domain configuration \cite{Santa-aho2026, Nadoum2020, Batista2014}. In the present work, EBSD was first used to identify grains with the crystallographic c-axis oriented normal to the sample surface, thereby establishing a crystallographic reference. Subsequently, Kerr microscopy was performed on the same grains. 

\begin{figure*}[t!]
	\centering
	\includegraphics[width=0.8\textwidth]{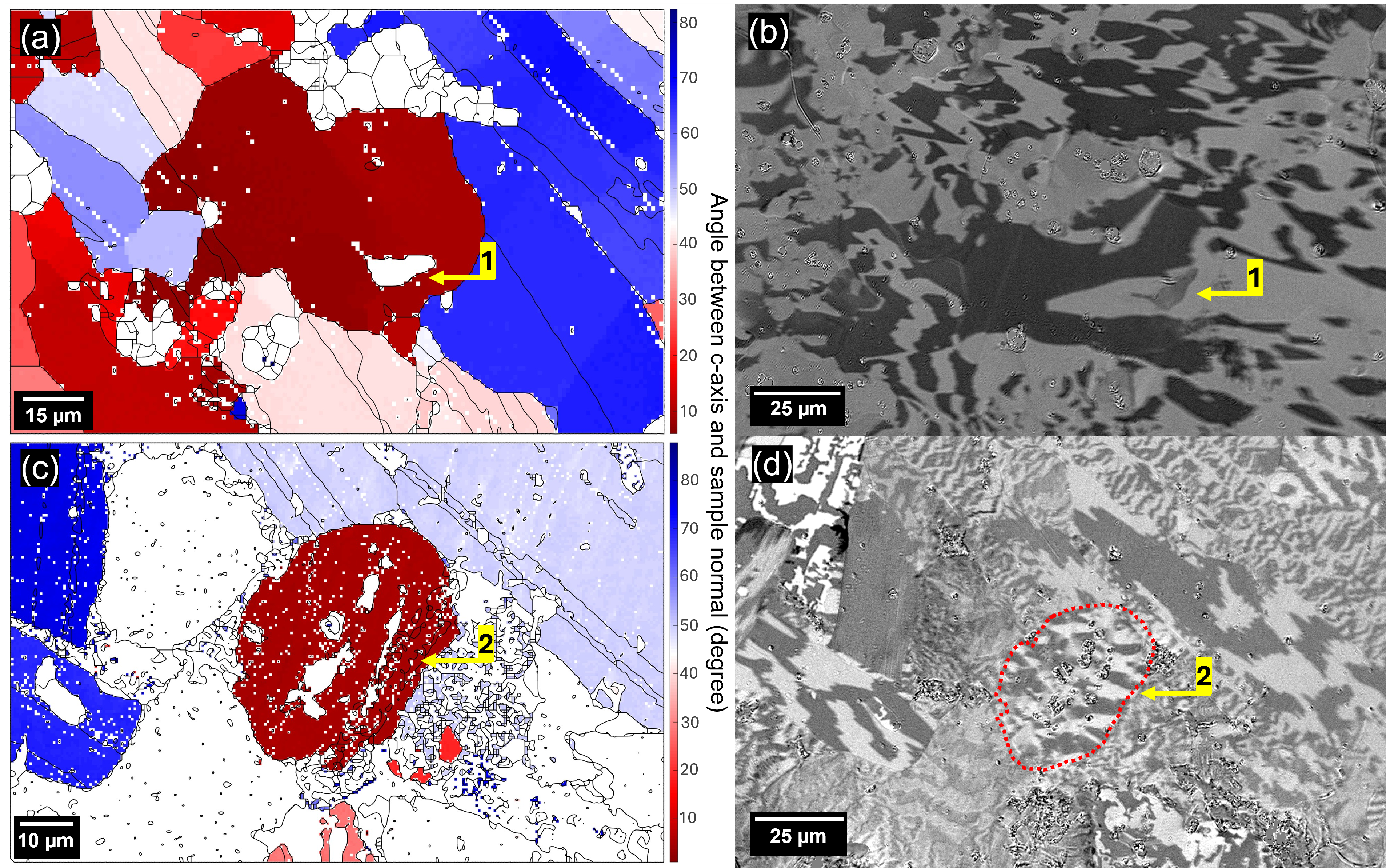}
	\caption{Correlative EBSD and Kerr microscopy analysis of V-doped YFe$_3$. (a) EBSD orientation map of the sample with 1 at. \% V, colored according to the angle between the crystallographic c-axis of the YFe$_3$ phase and the sample normal; grains in red correspond to c-axis aligned out of plane. The marked arrow indicates the microstructural feature reference used for comparison with the Kerr microscopy analysis; white areas indicate non-indexed regions (secondary phases or pores). (b) Corresponding Kerr micrograph from the selected grain in (a), showing exclusively transverse contrast, indicating in-plane magnetization. (c) EBSD orientation map of the sample with 3 at. \% V (2 at. \% V in YFe$_3$). (d) Corresponding Kerr micrograph from the selected grain in (c), exhibiting purely transverse contrast and a stripe-like magnetic domain pattern, characteristic of in-plane magnetization.}
	\label{fig:Fig10}
\end{figure*}

The EBSD orientation map shown in Fig.~\ref{fig:Fig10}(a), obtained from the as-heat-treated sample with an initial composition of 1 at. \% V, is colored according to the angle between the crystallographic c-axis and the sample normal. The grain in red corresponds to the region where the c-axis is oriented parallel to the out-of-plane direction of the image. This region was selected for subsequent Kerr microscopy analysis. Areas shown in white correspond to “no solution” in the YFe$_3$ phase indexing, indicating either another phase or pores. 

Panel (b) shows the corresponding Kerr micrograph acquired from the selected region, revealing transverse contrast. Because the transverse magneto-optical Kerr effect is sensitive to in-plane magnetization, the absence of polar Kerr contrast indicates that the magnetization is not aligned along the out-of-plane (c-axis) direction \cite{Soldatov2020, Soldatov2017}. Instead, the observed magnetic domain configuration is characteristic of in-plane magnetization. Such stripe domains are well known to indicate in-plane magnetization, whereas “cog-wheel” or flower-like domain patterns are typically associated with out-of-plane magnetization components, as reported by Khlopkov \textit{et al.} (2004) \cite{Khlopkov2004} and more recently by Choudhary \textit{et al.} (2024) \cite{Choudhary2022}. These referenced domain patterns, observed in benchmark systems such as Nd-Fe-B magnets, serve as reference configurations for magnetic domain analysis in the present study.

The analysis was also performed on the as-heat-treated sample with an initial composition of 3 at. \% V, for which EDS point analysis confirmed approximately 2 at. \% V incorporation within the YFe$_3$ grains. The EBSD orientation map shown in Fig.~\ref{fig:Fig10}(c), colored according to the angle between the c-axis and the sample normal, reveals a grain with its c-axis aligned out of plane. Panel (d) presents the corresponding Kerr image of this grain, which exhibits purely transverse contrast without any polar contribution. The magnetic domain pattern displays the characteristic stripe-like morphology observed in the lower V-content sample, consistent with magnetization confined to the basal plane.

The absence of a detectable polar Kerr contribution, even in grains with the $c$-axis oriented out of plane, indicates that easy-axis anisotropy does not develop upon V incorporation up to approximately 2~at.\%. This result is consistent with the intrinsic magnetic behavior of the YFe$_3$ phase. Van der Kraan \textit{et al.} (1975) \cite{VanderKraan1975} established that the easy magnetization direction in YFe3 lies within the basal plane, and this easy-plane anisotropy is further supported quantitatively by Bolyachkin \textit{et al.} (2017) \cite{bolyachkin2017}, who reported a negative magnetocrystalline anisotropy constant (K$_1$ = \SI{-1.1}{\mega\joule\per\metre\cubed}).

The observed in-plane magnetic domain structures are therefore consistent with the calculated concentration dependence of the MAE. The experimentally achieved V incorporation of up to approximately 2~at.\% lies well below the calculated concentration range of approximately 11--14~at.\% at which the MAE changes sign, explaining why the intrinsic easy-plane anisotropy of YFe$_3$ is retained in the experimentally accessible samples.

Attempts to experimentally synthesize YFe$_{2.33}$Ta$_{0.67}$ were also performed within the scope of this work. Under the present arc-melting and heat-treatment conditions, appreciable Ta incorporation into the YFe$_3$ phase was not achieved, as Ta preferentially formed Fe$_2$Ta together with YFe$_2$ and $\alpha$-Y. Alternative non-equilibrium synthesis approaches, such as rapid solidification by melt spinning, may provide a route to suppress competing secondary phases and promote Ta incorporation into the 1:3 phase matrix. Further details are provided in Section~III.B of the Supplementary Material.

\section{CONCLUSIONS}

    In this work, we investigated the effect of transition-metal substitution on the magnetic properties and magnetocrystalline anisotropy of YFe$_3$ using a multiscale computational framework comprising first-principles density functional theory, atomistic Monte Carlo spin-dynamics simulations and micromagnetic modelling, complemented by experimental characterization of V-alloyed YFe$_3$. While pristine YFe$_3$ exhibits easy-plane anisotropy, our results demonstrate that selective substitution at the Fe sublattice provides an effective route for inducing uniaxial anisotropy while preserving strong ferromagnetic order and a high Curie temperature.

    Among the investigated dopants, V, Nb, and Ta systematically enhance the magnetocrystalline anisotropy. While V induces a moderate increase in uniaxial anisotropy, Nb and Ta are particularly effective in stabilizing a uniaxial magnetic ground state. In particular, YFe$_{2.33}$Ta$_{0.67}$ exhibits a large magnetocrystalline anisotropy energy of 2.30~MJ/m$^3$ together with a Curie temperature above 500~K. Calculations performed at fixed lattice geometries demonstrate that the easy-plane-to-easy-axis transition is driven primarily by dopant-induced modifications of the electronic structure rather than by structural relaxation.

    Atom- and orbital-resolved analyses reveal that the enhanced anisotropy originates from dopant-induced modifications of the electronic structure near the Fermi level. Within the second-order perturbation framework, SOC-mediated virtual transitions involving Fe $3d$ and dopant $d$ states give rise to substantial positive contributions to the MAE. Although the Fe-3$b$ site retains a local tendency toward easy-plane anisotropy, this contribution is outweighed by positive contributions from the Fe-18$h$, dopant-6$c$, and Y sites, resulting in a robust uniaxial magnetic ground state.
    
    The calculated Heisenberg exchange interactions show that magnetism in all investigated compounds remains governed by the Fe sublattice. While V slightly weakens the Fe--Fe exchange interactions, Nb and Ta largely preserve the ferromagnetic exchange network. Consequently, atomistic Monte Carlo simulations predict Curie temperatures close to that of pristine YFe$_3$, demonstrating that the enhancement of anisotropy is achieved without compromising thermal magnetic stability. Among the investigated compositions, YFe$_{2.33}$Ta$_{0.67}$ exhibits the largest MAE while retaining a Curie temperature above 500~K. Micromagnetic simulations reveal a clear trade-off between coercivity and energy-product performance. YFe$_{2.33}$Ta$_{0.67}$ exhibits the highest coercivity and the strongest resistance to irreversible demagnetization across the investigated microstructures, whereas YFe$_{2.33}$V$_{0.67}$ achieves a comparable $(BH)_{\max}$ at 300~K and the largest calculated $(BH)_{\max}$ at 450~K. These results show that Ta substitution is advantageous when coercive stability is prioritized, whereas V substitution provides competitive energy-product performance.

    Experimentally, V-alloyed YFe$_3$ synthesized by arc melting and subsequent heat treatment retains the YFe$_3$ phase at low nominal V concentrations, whereas higher V contents promote the formation of secondary phases. Correlative EBSD and Kerr microscopy reveal predominantly in-plane magnetic domain structures in this low-V range, consistent with the calculated concentration dependence of the MAE, which changes sign between approximately 11 and 14~at.\% V. Although synthesis of the predicted Ta-rich composition remains challenging under the present processing conditions, YFe$_{2.33}$Ta$_{0.67}$ emerges from the present multiscale calculations as a promising target for future experimental investigation.
    
    Overall, the present work establishes Fe-site substitution, particularly with V, Nb, and Ta, as an effective strategy for transforming the intrinsic easy-plane anisotropy of YFe$_3$ into uniaxial anisotropy while preserving strong ferromagnetic order and high thermal stability. These results suggest that targeted Fe-site substitution may offer a promising route toward the development of 4$f$-free Fe-rich intermetallic permanent-magnet materials.
    
\subsection*{Acknowledgments}
    The authors acknowledge computational resources and support provided by the National Academic Infrastructure for Supercomputing in Sweden (NAISS) (projects NAISS 2026/1-54, NAISS 2025/1-38) at NSC (Linköping University) and PDC (KTH Royal Institute of Technology). They gratefully acknowledge Dr. I. Soldatov and Prof. Dr. R. Schäfer for providing the Kerr microscopy images, and J. Winkler for performing the EBSD measurements. The authors thank Georgia A. Marchant and Junaid Jami for helpful discussions. ChatGPT (OpenAI) assisted with English language editing, with all revisions reviewed and approved by the authors.

    Funded by the European Union (Grant agreement ID: 101135546). The views and opinions expressed are, however, those of the author(s) only and do not necessarily reflect those of the European Union or the European Health and Digital Executive Agency (HADEA). Neither the European Union nor the granting authority can be held responsible for them.\\

\subsection*{Competing interests}
The authors declare no competing interests.

\bibliography{references}

\end{document}